\documentclass[aps,pra,twocolumn]{revtex4-2}

\usepackage{amsmath}
\usepackage{amssymb}
\usepackage{bm}
\usepackage{graphicx}
\usepackage{mathtools}
\usepackage{braket}
\usepackage{braket}

\begin{document}

\title{Exact Analysis of a One-Dimensional Yang-Gaudin Model with Two-Body Loss}

\author{Ryutaro Katsuta}
\affiliation{Department of Electronic and Physical Systems, Waseda University, Tokyo 169-8555, Japan}

\author{Shun Uchino}
\affiliation{Department of Electronic and Physical Systems, Waseda University, Tokyo 169-8555, Japan}
\affiliation{Department of Materials Science, Waseda University, Tokyo 169-8555, Japan}

\date{\today}

\begin{abstract}
We show that the one-dimensional Yang–Gaudin model with two-body loss remains exactly solvable irrespective of whether constituent
particles are bosons or fermions. 
By relating the Liouvillian spectrum to the right eigenvalues of a non-Hermitian effective Hamiltonian
obtained by complexifying the interaction strength,
we derive a general expression for the initial particle-loss rate.
We then solve the two-body problem exactly and show that, in the bosonic singlet sector,
the effective Hamiltonian has real right eigenvalues and the master equation admits steady-state solutions
For many-body systems with three or more particles, we further show that
dissipation reverses which spin configurations are most stable: in bosonic
systems it favors antiferromagnetic-like configurations over ferromagnetic-like ones,
whereas in fermionic systems it favors ferromagnetic-like configurations over antiferromagnetic-like ones.
\end{abstract}

\maketitle

\newcommand{\hpsi}{\hat{\psi}}
\section{Introduction}
In realistic experimental settings, quantum systems are inevitably coupled to an environment. 
Such system–environment coupling induces dissipation and decoherence, and 
the resulting  dynamics cannot, in general, be described by the unitary time evolution of isolated systems. 
Therefore,  a description within the framework of open quantum systems is essential for developing a theoretical understanding of experimentally realizable quantum many-body systems~\cite{re:open_quan}.
Under the Markov approximation, open-system dynamics are govened by the Gorini–Kossakowski–Sudarshan–Lindblad (GKSL) quantum master equation~\cite{re:open_quan,re:master_eq_1,re:master_eq_2}.
A central theoretical challenge is to understand interacting open quantum systems in settings where
analytical control is possible.

One important route toward such an understanding is provided by exactly solvable models.
By solving the eigenvalue problem of the Liouvillian, one can analyze the time evolution governed by the master equation, and several exactly solvable examples are known~\cite{re:kakaimodel_1,re:kakaimodel_2,re:kakaimodel_3,re:kakaimodel_4,re:kakaimodel_5,re:kakaimodel_6,re:kakaimodel_7,re:kakaimodel_8,re:kakaimodel_9,re:kakaimodel_10,re:kakaimodel_11,re:kakaimodel_12}. 
Even so, extending integrability from closed to open systems remains highly nontrivial;
the integrability of the Hamiltonian part does not by itself guarantee that 
 the corresponding Liouvillian can be analyzed exactly.
In this context, one-dimensional gases with contact interactions are particlularly attractive.
Their closed-system counterparts include the Lieb-Liniger model for spinless bosons~\cite{re:LiebLiniger}
and the fermionic Yang-Gaudin model~\cite{re:YangGaudin_1,re:YangGaudin_2},
both of which are paradigmatic Bethe-ansatz-solvable systems.
Moreover, two-component Bose gases with approximate U(2) symmetry have been realized in ${}^{87}\mathrm{Rb}$~\cite{re:U2BEC_1} and ${}^{23}\mathrm{Na}$~\cite{re:U2BEC_2}. 
In one dimension, such ssytems are equivalent to  spin-$1/2$ bosons and are likewise solvable by
the Bethe ansatz~\cite{re:B_YangGaudin_1,re:B_YangGaudin_2}, which further motivates
the study of spin-$1/2$ Yang-Gaudin-type models~\cite{re:ABA_BYG}.

Dissipation in one-dimensional integrable gases is known to generate qualitatively 
new physics rather than merely destroying coherent dynamics.
In the Lieb-Liniger setting, strong inelastic interactions can suppress particle loss
and effectively realize a Tonks-Giradeau gas~\cite{re:dis_Tonks_ex,re:dis_Tonks_the},
providing a notable example of dissipation-induced behavior. 
Another interesting example is the Fermi–Hubbard model
with two-body loss that can exhibit changes in the spin configurations favored by the dynamics~\cite{re:F_hubbard_1,re:kakaimodel_12}. 
These developments naturally raise the following question: can the one-dimensional spin-$1/2$
Yang-Gaudin model with two-body loss still be treated exactly, and if so, how does dissipation modify
its stable spin configurations?

In this work, we answer this question by showing that  the one-dimensional spin-$1/2$ Yang–Gaudin model with two-body loss remains exactly solvable independent of particle statistics. 
We first relate the Liouvillian spectrum to the eigenvalues of a non-Hermitian effective Hamiltonian 
obtained by complexifying the interaction strength, and derive a general expression for the initial particle-loss rate.
We then solve the two-body problem exactly and show that, in the bosonic singlet sector, the effective Hamiltonian
has real right eigenvalues and the master equation admits steady-state solutions.
For many-body systems with three or more particles, we further show that dissipation reverses the 
preferred spin configuration: in bosonic systems it favors antiferromagnetic-like configurations over
ferromagnetic ones, whereas in fermionic systems it favors ferromagnetic-like configurations over antiferromagnetic-like ones.

\section{The model}
We consider a one-dimensional spin-1/2 Yang-Gaudin model with two body loss.
In this section, we introduce the master equation describing the system and discuss the significance of 
the eigenvalues of the non-Hermitian effective Hamiltonian. In particular, Sec.~\ref{subsec_2_1} explains the relation between the eigenvalues of the Liouvillian and those of the non-Hermitian operator, Sec.~\ref{subsec_2_2} 
summarizes the Bethe ansatz formulation, and
Sec.~\ref{subsec_2_3} discusses
the relation between the eigenvalues of the non-Hermitian operator and the particle-loss rate.

\subsection{Master equation}\label{subsec_2_1}
We consider an open quantum many-body system with two body loss, described by the 
Gorini--Kossakowski--Sudarshan--Lindblad (GKSL) master equation~\cite{re:open_quan,re:master_eq_1,re:master_eq_2}
\begin{equation}\label{eq:master_equation_1} 
\begin{split}
    \frac{d\rho}{dt}
    =&-\frac{i}{\hbar}[\hat{H},\rho]\\
    &+\frac{\gamma}{2}\sum_{\sigma_\alpha,\sigma_\beta=\uparrow,\downarrow}\int dx[2\hpsi_{\sigma_\alpha}(x)\hpsi_{\sigma_\beta}(x)\rho
    \hpsi^{\dagger}_{\sigma_\beta}(x)
    \hpsi^{\dagger}_{\sigma_\alpha}(x)\\
    &-\lbrace\hpsi^{\dagger}_{\sigma_\beta}(x)
    \hpsi^{\dagger}_{\sigma_\alpha}(x)
    \hpsi_{\sigma_\alpha}(x)\hpsi_{\sigma_\beta}(x)
    ,\rho\rbrace]
    \equiv \mathcal{L}\rho.
\end{split}
\end{equation}
Here, $\gamma>0$, $\rho(t)$ denotes the density matrix at time $t$, and $\hat{\psi}_\sigma(x)$ and $\hat{\psi}_\sigma^{\dagger}(x)$ denote the annihilation and creation operators at position $x$, respectively. The commutator and anticommutator are denoted by $[\cdots,\cdots]$ and $\lbrace\cdots,\cdots\rbrace$, respectively. We consider a system in which the particle loss rates for parallel and antiparallel spins are equal. When $\rho$ represents a steady state, it satisfies
\begin{equation}
    \mathcal{L}\rho
    =\frac{d\rho}{dt}=\bm{0},
\end{equation}
and therefore the steady state corresponds to the eigenoperator of $\mathcal{L}$ with eigenvalue $0$. 
Hence, by analyzing the eigenvalues of the Liouvillian, one can determine whether a stable state exists.

The Hamiltonian of the Yang–Gaudin model is given by~\cite{re:YangGaudin_1}
\begin{equation}
\begin{split}
    \hat{H}
    =&-\frac{\hbar^2}{2m}
    \sum_{\sigma=\uparrow,\downarrow}
    \int dx\; \hpsi^\dagger_\sigma (x) \frac{\partial^2}{\partial x^2}\hpsi_\sigma(x)\\
    &+2c\sum_{\sigma_\alpha,\sigma_\beta=\uparrow,\downarrow}
    \int dx\; \hpsi^{\dagger}_{\sigma_\alpha}(x)
    \hpsi^{\dagger}_{\sigma_\beta}(x)\hpsi_{\sigma_\beta}(x)\hpsi_{\sigma_\alpha}(x),
\end{split}
\end{equation}
where $m$ is the mass of a single particle. Here we introduce
\begin{equation}
\begin{split}
    &H_{\mathrm{eff}}\\
    &\equiv\hat{H}
    -\frac{i\hbar\gamma}{2}\sum_{\sigma_\alpha,\sigma_\beta
    =\uparrow,\downarrow}
    \int dx\; \hpsi^{\dagger}_{\sigma_\alpha}(x)
    \hpsi^{\dagger}_{\sigma_\beta}(x)\hpsi_{\sigma_\beta}(x)\hpsi_{\sigma_\alpha}(x)\\
    &=-\frac{\hbar^2}{2m}
    \sum_{\sigma=\uparrow,\downarrow}
    \int dx\; \hpsi^\dagger_\sigma (x) \frac{\partial^2}{\partial x^2}
    \hpsi_\sigma(x)
    +2\left(c-\frac{i\hbar\gamma}{4}\right)\\
    &\;\;\;\;\times\sum_{\sigma_\alpha,\sigma_\beta
    =\uparrow,\downarrow}
    \int dx\; \hpsi^{\dagger}_{\sigma_\alpha}(x)
    \hpsi^{\dagger}_{\sigma_\beta}(x)\hpsi_{\sigma_\beta}(x)\hpsi_{\sigma_\alpha}(x),
\end{split}
\end{equation}
and define
\begin{equation}
    \label{eq:def_K_A}
    \mathcal{K}\rho\equiv\frac{1}{i\hbar}(H_{\mathrm{eff}}\rho-\rho H_{\mathrm{eff}}^\dagger),
\end{equation}
\begin{equation}
    \label{eq:def_A}
    \mathcal{A}\rho\equiv\gamma
    \sum_{\sigma_\alpha,\sigma_\beta
    =\uparrow,\downarrow}\int dx\;\hpsi_{\sigma_\alpha}(x)\hpsi_{\sigma_\beta}(x)\rho
    \hpsi^{\dagger}_{\sigma_\beta}(x)
    \hpsi^{\dagger}_{\sigma_\alpha}(x),
\end{equation}
so that Eq. \eqref{eq:master_equation_1} can be rewritten as
\begin{equation}\label{eq:master_equation_2} 
   \mathcal{L}\rho=\mathcal{K}\rho+\mathcal{A}\rho.
\end{equation}
We consider a system with a finite maximum particle number $N$.
Let $n = 0, 1, \cdots, N$ denote the particle number, and let $a^*$ denote the complex conjugate of a complex number $a$. We assume that there exist complete sets ${\lvert r_j^n \rangle}$ and ${\lvert q_j^n \rangle}$ in the Hilbert space $\mathcal{H}$ satisfying
\begin{equation}
    H_{\mathrm{eff}}\Ket{r_j^n}=\varepsilon_j^{(n)}\Ket{r_j^n},\;\;\;
    H_{\mathrm{eff}}^\dagger\Ket{q_j^n}=\varepsilon_j^{(n)*}\Ket{q_j^n},
\end{equation}
\begin{equation}
    \Braket{q_k^n|r_j^m}=\delta_{k,j}\delta_{n,m},
\end{equation}
\begin{equation}
    \label{eq:comp_1}
    \sum_{n=0}^N\sum_{j=1}^{\infty}
    \Ket{r_j^n}\Bra{q_j^n}=\mathbf{I}_{\mathcal{H}},
\end{equation}
where $\mathbf{I}_{\mathcal{H}}$ denotes the identity operator on $\mathcal{H}$. In addition, we impose the following mathematical assumptions:
 (a) we assume that $\varsigma(\mathcal{K})=\varsigma_P(\mathcal{K})$, where $\varsigma(T)$ and $\varsigma_P(T)$ denote the spectrum and the set of eigenvalues of operator $T$;
 (b) we assume that $\mathcal{A}$ is bounded.
In this case, the set of eigenvalues of $\mathcal{K}$, denoted by $\varsigma_P(\mathcal{K})$, is given by
\begin{equation}\label{eigenvalue_K}
    \varsigma_P(\mathcal{K})
    =\left\lbrace
    \frac{1}{i\hbar}\left(
    \varepsilon_{j}^{(n)}-\varepsilon_{k}^{(l)*}
    \right)
    \right\rbrace_{\substack{n,l=0,1,2,\cdots,N\\j,k=1,2,\cdots}},
\end{equation}
then as shown in Appendix A,
\begin{equation}\label{inclusion}
    \varsigma_P(\mathcal{L})\subset
    \varsigma_P(\mathcal{K}).
\end{equation}
From Eqs.~\eqref{eigenvalue_K} and \eqref{inclusion}, a necessary condition for the system described by Eq.~\eqref{eq:master_equation_1} to possess a steady state is that there exist eigenvalues $\varepsilon_{j}^{(n)}$ and $\varepsilon_{k}^{(l)}$ of $H_{\mathrm{eff}}$ satisfying
\begin{equation}
    \label{eq:cond_stedy}
    \varepsilon_{j}^{(n)}=\varepsilon_{k}^{(l)*}.
\end{equation}
Therefore, by analyzing the eigenvalues of $H_{\mathrm{eff}}$, one can infer the possible eigenvalues of the Liouvillian.

\subsection{Bethe ansatz}\label{subsec_2_2}
Here, we obtain the right eigenstates of $H_{\mathrm{eff}}$ using the Bethe ansatz.
To this end, we consider a one-dimensional gas in a box of length $L$ with periodic boundary conditions. When the particle number is $n$, $H_{\mathrm{eff}}$ is given by
\begin{equation}
    H_{\mathrm{eff}}
    =-\frac{\hbar^2}{2m}\sum_{j=1}^n \frac{\partial^2}{\partial x_j^2}
    +2\left(
    c-\frac{i\hbar\gamma}{4}
    \right)\sum_{i<j}\delta(x_i-x_j).
\end{equation}

Let $V\cong\mathbb{C}^2$ be the single-particle spin space for spin-$1/2$. Then the $n$-particle spin space is given by $V_1\otimes\cdots\otimes V_n$. We denote an orthonormal basis of this space by $\ket{\sigma_1,\cdots,\sigma_n}$, where $\sigma_j(j=1,\cdots,n)$ takes either $\uparrow$ or $\downarrow$.
According to the Bethe ansatz, the right eigenstate in the region $0\leq x_{Q_1}\leq\cdots\leq x_{Q_n}$ is given by \cite{re:YangGaudin_1}
\begin{equation}\label{eq:bethe_wavefunction}
\begin{split}
    &\psi_R(x_1,\cdots,x_n)\\
    &=\sum_{\sigma_1,\cdots,\sigma_n}
    \psi_R(x_1,\cdots,r_n,\sigma_1,
    \cdots,\sigma_n)\ket{\sigma_1,\cdots,\sigma_n},\\
    &\psi_R(x_1,\cdots,x_n,\sigma_1,
    \cdots,\sigma_n)\\
    &=\sum_{P\in\mathfrak{P}^n}\mathfrak{A}
    A_{\sigma_{Q_1}\cdots\sigma_{Q_n}}(k_{P_1},\cdots,k_{P_n})\exp\left[
    i\sum_{j=1}k_{P_j}x_{Q_j}
    \right],\\
\end{split}
\end{equation}
where $\mathfrak{P}^n$ denotes the set of all permutations of $\lbrace1,\cdots,n\rbrace$, and $Q\in\mathfrak{P}^n$. The factor $\mathfrak{A}$ is given by
\begin{equation}
    \mathfrak{A}
    =\begin{cases}
        1,&(\mathrm{Boson})\\
        \mathrm{sgn}(PQ),&(\mathrm{Fermion})
    \end{cases},
\end{equation}
where $\mathrm{sgn}(P)$ denotes the signature of the permutation $P$. An explicit form of the right eigenstate is given in Appendix B, Eq.~\eqref{eq:bethe_wavefunction_2}.

When the number of $\downarrow$ spins is $M$, the Bethe equations for the quasimomenta$\;k_1,\cdots,k_n$  and the spin rapidities $l_1,\cdots,l_M$ are, for bosons \cite{re:B_YangGaudin_1,re:B_YangGaudin_2}
\begin{equation}\label{eq_bethe_boson_1}
    e^{ik_jL}
    =-\prod_{l=1}^n\frac{k_j-k_l+ic'}{k_j-k_l-ic'}
    \prod_{\beta=1}^M\frac{k_j-l_\beta-ic'/2}{k_j-l_\beta+ic'/2},
\end{equation}
\begin{equation}\label{eq_bethe_boson_2}
    \prod_{j=1}^n\frac{l_\alpha-k_j-ic'/2}{l_\alpha-k_j+ic'/2}
    =-\prod_{\beta=1}^M\frac{l_\alpha-l_\beta-ic'}{l_\alpha-l_\beta+ic'},
\end{equation}
and for fermions \cite{re:YangGaudin_1,re:YangGaudin_2}
\begin{equation}\label{eq_bethe_fermion_1}
    e^{ik_jL}
    =\prod_{\alpha=1}^M\frac{k_j-l_\alpha+ic'/2}{k_j-l_\alpha-ic'/2},
\end{equation}
\begin{equation}
    \prod_{j=1}^n\frac{l_\alpha-k_j+ic'/2}{l_\alpha-k_j-ic'/2}
    =-\prod_{\beta=1}^M\frac{l_\alpha-l_\beta+ic'}{l_\alpha-l_\beta-ic'},
    \label{eq_bethe_fermion_2}
\end{equation}
where we define $c'\equiv2m(c-i\hbar\gamma/4)/\hbar^2$.
We denote the total spin operator by $\hat{\bm{S}}$. The right eigenstates obtained from the Bethe equations correspond to highest-weight states for a given eigenvalue of $\hat{\bm{S}}^2$ (see Appendix B). Since the lowering operator $S^-$ commutes with the Hamiltonian, acting with $S^-$ generates right eigenstates with different spin configurations that share the same right eigenvalue.
From Eqs.~\eqref{eq_bethe_boson_1} and \eqref{eq_bethe_boson_2}, for bosons we obtain
\begin{equation}
\label{eq:Betheeq_qu1}
    k_jL=2\pi I_j
    -2\sum_{l=1}^n\tan^{-1}\frac{k_j-k_l}{c'}
    +2\sum_{\beta=1}^M\tan^{-1}\frac{2k_j-2l_\beta}{c'},
\end{equation}
\begin{equation}
\label{eq:Betheeq_qu2}
    \pi J_\alpha
    =\sum_{j=1}^n\tan^{-1}
    \frac{2l_\alpha-2k_j}{c'}
    -\sum_{\beta=1}^M\tan^{-1}\frac{l_\alpha-l_\beta}{c'},
\end{equation}
where $\tan^{-1}(x)$ denotes the principal value, and $I_j$ and $J_\alpha$ are integers (half-integers) when $n-M$ is odd (even). From Eqs.~\eqref{eq_bethe_fermion_1} and \eqref{eq_bethe_fermion_2}, for fermions we obtain
\begin{equation}
\label{eq:Betheeq_qu3}
    k_jL
    =2\pi I_j
    -2 \sum_{\alpha=1}^M \tan^{-1}\frac{2k_j-2l_\alpha}{c'},
\end{equation}
\begin{equation}
\label{eq:Betheeq_qu4}
    \pi J_\alpha
    =-\sum_{j=1}^n\tan^{-1}\frac{2l_\alpha-2k_j}{c'}
    +\sum_{\beta=1}^M \tan^{-1}\frac{l_\alpha-l_\beta}{c'},
\end{equation}
where $I_j$ are integers (half-integers) when $M$ is even (odd), and $J_\alpha$ are integers (half-integers) when $M-n$ is odd (even). The right eigenvalues are given by
\begin{equation}
    E=\frac{\hbar^2}{2m}\sum_{j=1}^n k_j^2.
\end{equation}

In the following, we assume that the quasi-momenta $k_1,\cdots,k_n$ take mutually distinct values, that the Bethe ansatz wave function given by Eq.~\eqref{eq:bethe_wavefunction} forms a complete set, and that the corresponding left eigenstates exist. Under these assumptions, we proceed with the analysis.

\subsection{Loss rate}\label{subsec_2_3}
In Bose-Einstein condensation, a relation between the particle-loss rate and the right eigenvalues of $H_{\mathrm{eff}}$ is known~\cite{re:dis_Tonks_the}. Here we show that the same relation holds more generally.

The initial particle-loss rate can be obtained from the master equation. 
The time derivative of the expectation value of the total-number operator $\hat{N}=\sum_{\sigma}\int dx\hat{\psi}_{\sigma}^\dagger(x)\hat{\psi}_{\sigma}(x)$ is
\begin{equation}
\begin{split}
    &\frac{d\Braket{\hat{N}}}{dt}
    =\frac{d}{dt}\mathrm{Tr}\left[\hat{N}\rho\right]\\
    &=-2\gamma\sum_{\sigma,\sigma'}
    \int dx\Braket{\hat{\psi}^\dagger_\sigma(x)
    \hat{\psi}^\dagger_{\sigma'}(x)
    \hat{\psi}_{\sigma'}(x)
    \hat{\psi}_\sigma(x)},
\end{split}
\end{equation}
which is expressed in terms of the local two-body correlation function.
From Eqs.~\eqref{eq:master_equation_1} and \eqref{eq:master_equation_2}, we obtain
\begin{equation}
\begin{split}
    &\frac{d\mathrm{Tr}[\rho]}{dt}
    =\mathrm{Tr}\left[\frac{1}{i\hbar}\left(
    H_{\mathrm{eff}}\rho-\rho H^\dagger_{\mathrm{eff}}\right)
    +\mathcal{A}\rho
    \right]\\
    &=\frac{1}{i\hbar}\mathrm{Tr}\left[
    H_{\mathrm{eff}}\rho-\rho H^\dagger_{\mathrm{eff}}\right]
    -\frac{1}{2}\frac{d\Braket{\hat{N}}}{dt}.
\end{split}
\end{equation}
On the other hand,
\begin{equation}
\begin{split}
    &\frac{d\mathrm{Tr}[\rho]}{dt}\\
    &=-\frac{i}{\hbar}\mathrm{Tr}\left[[\hat{H},\rho]\right]\\
    &\;\;\;\;\;+\frac{\gamma}{2}\sum_{\sigma_\alpha,\sigma_\beta=\uparrow,\downarrow}\int dx\mathrm{Tr}
    [2\hpsi_{\sigma_\alpha}(x)\hpsi_{\sigma_\beta}(x)\rho
    \hpsi^{\dagger}_{\sigma_\beta}(x)
    \hpsi^{\dagger}_{\sigma_\alpha}(x)\\
    &\;\;\;\;\;-\lbrace\hpsi^{\dagger}_{\sigma_\beta}(x)
    \hpsi^{\dagger}_{\sigma_\alpha}(x)
    \hpsi_{\sigma_\alpha}(x)\hpsi_{\sigma_\beta}(x)
    ,\rho\rbrace]\\
    &=0.
\end{split}
\end{equation}
Therefore,
\begin{equation}
    \frac{d\Braket{\hat{N}}}{dt}
    =\frac{2}{i\hbar}\mathrm{Tr}\left[
    H_{\mathrm{eff}}\rho-\rho H^\dagger_{\mathrm{eff}}\right].
\end{equation}
If
\begin{equation}
    \rho(0)=\sum_{n.n',j,j'}
    c_{j,j'}^{(n,n')}
    \ket{r_j^n}\bra{r_{j'}^{n'}},
\end{equation}
then
\begin{equation}\label{eq:relarion_loss_E_1}
\begin{split}
    &\left.\frac{d\Braket{\hat{N}}}{dt}\right|_{t=0}
    =\frac{2}{i\hbar}\mathrm{Tr}\left[
    H_{\mathrm{eff}}\rho(0)-\rho(0) H^\dagger_{\mathrm{eff}}\right]\\
    &=\frac{2}{i\hbar}
    \mathrm{Tr}\left[
    \sum_{n.n',j,j'}
    \left(
    \varepsilon_{j}^{(n)}-\varepsilon_{j'}^{(n')*}
    \right)
    c_{j,j'}^{(n,n')}
    \ket{r_j^n}\bra{r_{j'}^{n'}}
    \right].
\end{split}
\end{equation}
In particular, if
\begin{equation}
    \rho(0)=
    \ket{r_j^n}\bra{r_{j}^{n}},
\end{equation}
then, using $\Braket{r_j^n|r_j^n}=1$, we obtain
\begin{equation}
    \label{eq:relarion_loss_E_2}
    \left.\frac{d\Braket{\hat{N}}}{dt}\right|_{t=0}
    =\frac{4\mathrm{Im}\left(
    \varepsilon_{j}^{(n)}
    \right)}{\hbar}.
\end{equation}
Therefore, when the initial density matrix  is  a pure state corresponding to a right eigenstate of $H_{\mathrm{eff}}$, the particle loss rate at the initial time is determined by the imaginary part of the corresponding right eigenvalue of $H_{\mathrm{eff}}$. Since states with a larger magnitude of the particle-loss rates can be regarded as more unstable, the stability of the corresponding right eigenstate can be examined by evaluating the imaginary part of its right eigenvalue.

\section{The two-body problem}
In this section, we consider the case $N=2$. The right eigenstates obtained from the Bethe ansatz are highest-weight states: for $M=0$, one obtains triplet states, while for $M=1$, one obtains singlet states. The Hilbert space $\mathcal{H}$ describing the system is decomposed as the direct sum
\begin{equation}
    \mathcal{H}
    =\mathcal{H}_0\oplus \mathcal{H}_1\oplus
    \cdots \oplus \mathcal{H}_N,
\end{equation}
where $\mathcal{H}_n$ denotes the $n$-particle Hilbert space. Furthermore, the $n$-particle Hilbert space can be written as a tensor product of coordinate and spin parts:
\begin{equation}
    \label{eq:tens_n}
    \mathcal{H}_n
    =\mathcal{H}_{n,\mathrm{coord}}
    \otimes
    \mathcal{H}_{n,\mathrm{spin}}.
\end{equation}
In the two-particle case, the states
\begin{equation}
    \Ket{\uparrow\uparrow},\;
    \frac{1}{\sqrt{2}}(\Ket{\uparrow\downarrow}+\Ket{\downarrow\uparrow}),\;
    \Ket{\downarrow\downarrow},\;
    \frac{1}{\sqrt{2}}
    (\Ket{\uparrow\downarrow}-\Ket{\downarrow\uparrow}),
\end{equation}
form a basis of $\mathcal{H}_{2,\mathrm{spin}}$. Therefore, it can be decomposed as
\begin{equation}
    \label{eq:sum_spin}
    \mathcal{H}_{2,\mathrm{spin}}
    =\mathcal{H_{\mathrm{triplet}}}
    \oplus
    \mathcal{H_{\mathrm{singlet}}}.
\end{equation}
From Eqs.~\eqref{eq:tens_n} and \eqref{eq:sum_spin}, we obtain
\begin{equation}
    \mathcal{H}_{2}
    \cong
    (\mathcal{H}_{2,\mathrm{coord}}\otimes
    \mathcal{H_{\mathrm{triplet}}})
    \oplus
    (\mathcal{H}_{2,\mathrm{coord}}\otimes
    \mathcal{H_{\mathrm{singlet}}}).
\end{equation}
We consider situations in which the initial state is prepared in either a triplet or a singlet state, and investigate its stability. Since $\mathcal{K}$ preserves the particle number and $\mathcal{A}$ lowers the particle number by two (see Appendix A) without changing the number of $\downarrow$ spins, it suffices to 
restrict our attention to the spaces
\begin{equation}
    \mathcal{H}_0
    \oplus
    (\mathcal{H}_{2,\mathrm{coord}}\otimes
    \mathcal{H_{\mathrm{triplet}}}),
\end{equation}
or
\begin{equation}
    \mathcal{H}_0
    \oplus
    (\mathcal{H}_{2,\mathrm{coord}}\otimes
    \mathcal{H_{\mathrm{singlet}}}).
\end{equation}

\subsection{Singlet solutions of the bosonic Yang-Gaudin model}
\label{subsection:3_1}
Now we consider the two-body bosonic case in the singlet sector. 
We show that the right eigenvalues of $H_{\mathrm{eff}}$ are real. From the Bethe equations Eqs.~\eqref{eq_bethe_boson_1} and \eqref{eq_bethe_boson_2}, we obtain
\begin{equation}\label{eq:2body_bethe_AntiBoson_1}
     e^{ik_1L}
    =\frac{k_1-k_2+ic'}{k_1-k_2-ic'}
    \frac{k_1-l-ic'/2}{k_1-l+ic'/2},
\end{equation}
\begin{equation}\label{eq:2body_bethe_AntiBoson_2}
     e^{ik_2L}
    =\frac{k_2-k_1+ic'}{k_2-k_1-ic'}
    \frac{k_2-l-ic'/2}{k_2-l+ic'/2},
\end{equation}
\begin{equation}\label{eq:2body_bethe_AntiBoson_3}
    \frac{l-k_1-ic'/2}{l-k_1+ic'/2}
    \frac{l-k_2-ic'/2}{l-k_2+ic'/2}
    =1.
\end{equation}
Multiplying Eqs.~\eqref{eq:2body_bethe_AntiBoson_1} and \eqref{eq:2body_bethe_AntiBoson_2}, and using Eq.~\eqref{eq:2body_bethe_AntiBoson_3}, we obtain
\begin{equation}
\label{2body_AntiBoson_1}
\begin{split}
    &e^{i(k_1+k_2)L}=1\\
    &\Leftrightarrow
    \exists n\in\mathbb{Z},
    k_1+k_2
    =\frac{2n\pi}{L}.
\end{split}
\end{equation}

We assume $l-k_1+ic'/2\neq0$ and $l-k_2+ic'/2\neq0$. From Eqs.~\eqref{eq:2body_bethe_AntiBoson_3} and \eqref{2body_AntiBoson_1}, we obtain
\begin{equation}\label{2body_AntiBoson_2}
\begin{split}
    \frac{l-k_1-ic'/2}{l-k_1+ic'/2}
    \frac{l-k_2-ic'/2}{l-k_2+ic'/2}
    =1
    \Leftrightarrow
    l=\frac{n\pi}{L}.
\end{split}
\end{equation}
Substituting Eqs.~\eqref{2body_AntiBoson_1} and \eqref{2body_AntiBoson_2} into Eq.~\eqref{eq:2body_bethe_AntiBoson_1}, we find
\begin{equation}
\begin{split}
    &e^{ik_1L}
    =\frac{k_1+k_1-\frac{2n\pi}{L}+ic'}{k_1+k_1-\frac{2n\pi}{L}-ic'}
    \frac{k_1-\frac{n\pi}{L}-ic'/2}{k_1-\frac{n\pi}{L}+ic'/2}
    =1\\
    &\Leftrightarrow
    \exists n'\in\mathbb{Z},
    k_1=\frac{2n'\pi}{L}.
\end{split}
\end{equation}
Summarizing the above, for arbitrary complex $c'$, we obtain
\begin{equation}
\label{eq:prop_b_YG}
\begin{split}
    &\begin{dcases}
    e^{ik_1L}
    =\frac{k_1-k_2+ic'}{k_1-k_2-ic'}
    \frac{k_1-l-ic'/2}{k_1-l+ic'/2}\\
    e^{ik_2L}
    =\frac{k_2-k_1+ic'}{k_2-k_1-ic'}
    \frac{k_2-l-ic'/2}{k_2-l+ic'/2}\\
    \frac{l-k_1-ic'/2}{l-k_1+ic'/2}
    \frac{l-k_2-ic'/2}{l-k_2+ic'/2}
    =1
    \end{dcases}\\
    &\Leftrightarrow
    \exists n,n'\in\mathbb{Z},
    \begin{dcases}
        k_1=\frac{2n'\pi}{L}\\
        k_2
        =-k_1+\frac{2n\pi}{L}\\
        l=\frac{n\pi}{L}
    \end{dcases}.
\end{split}
\end{equation}
The corresponding right eigenvalue of $H_{\mathrm{eff}}$ is then given by
\begin{equation}
    E=\frac{\hbar^2}{2m}(k_1^2+k_2^2)
    =\frac{2\hbar^2(n'^2+(n-n')^2)\pi^2}{mL^2}
\end{equation}
which is a non-negative real number. This implies that even in the attractive case ($c<0$), there are no bound states, i.e., no string solutions. Moreover, since $H_{\mathrm{eff}}$ has only real right eigenvalues in this sector,  Eq.~\eqref{eq:relarion_loss_E_2} shows that the initial particle-loss rate vanishes for a pure state $\ket{r_j^2}\bra{r_{j}^{2}}$ constructed from the right eigenstates of $H_{\mathrm{eff}}$.

\subsection{Steady states of the bosonic Yang-Gaudin model}
\label{subsection:3_2}
We next provide explicit examples of eigenoperators of the Liouvillian with eigenvalue zero, using the right eigenstates of $H_{\mathrm{eff}}$ in the singlet sector of the two-body bosonic Yang–Gaudin model.

From Eq.~\eqref{eq:A_wave} in Appendix A, we have
\begin{equation}
\begin{split}
    &\mathcal{A}\Ket{r_j^2}\bra{r_{j'}^2}\\
    &=2\gamma
    \sum_{\sigma,\sigma'}
    \int dx
    \phi_{r_j^2}(x,x,\sigma,\sigma')\phi^*_{r_{j'}^2}(x,x,\sigma,\sigma')
    \Ket{0}\Bra{0},
\end{split}
\end{equation}
where $\phi_{r_j^2}(x_1,x_2,\sigma,\sigma')$ and $\phi_{r_{j'}^2}(x_1,x_2,\sigma,\sigma')$ are the wave functions corresponding to $\ket{r_j^2}$ and $\ket{r_{j'}^2}$, respectively, as given in Eq.~\eqref{eq:bethe_wavefunction}.
In the singlet case, Eq.~\eqref{eq:bethe_wavefunction_2} in Appendix B
yields
\begin{equation}
    \psi(x,x,\sigma_1,\sigma_2)=0,
\end{equation}
and therefore
\begin{equation}
    \mathcal{A}\Ket{r_j^2}\bra{r_{j'}^2}
    =0.
\end{equation}
Since the corresponding right eigenvalues $\varepsilon_j^{(2)}$   and $\varepsilon_j^{(2)}$ are real, it follows that whenever $\varepsilon_j^{(2)}=\varepsilon_{j'}^{(2)}$,
\begin{equation}
    \mathcal{L}\Ket{r_j^2}
    \Bra{r_{j'}^2}=0.
\end{equation}
Therefore, $\Ket{r_j^2}\Bra{r_{j'}^2}$ is a steady state. As a simple example, the pure state $\Ket{r_j^2}\Bra{r_j^2}$ is a steady state.
In this case, Eq.~\eqref{eq:relarion_loss_E_1} gives
\begin{equation}
\left.d\Braket{\hat{N}}/dt\right|_{t=0}=0
\end{equation}
when $\rho(0)=\Ket{r_j^2}\bra{r_{j}^2}$.

As will be shown in Secs.~\ref{subsection:3_3} and \ref{subsection:3_4}, 
the right eigenvalues of $H_{\mathrm{eff}}$ become complex 
in the bosonic Yang–Gaudin model in the triplet sector and 
in the fermionic Yang–Gaudin model in the singlet sector when dissipation is present. According to Eq.~\eqref{eq:relarion_loss_E_2}, the imaginary part of the right eigenvalues of $H_{\mathrm{eff}}$ determines  the initial particle-loss rate, and is therefore expected to be negative.
Under the assumption that no complex-conjugate partner exists for a given right eigenvalue,  Eq.~\eqref{eq:cond_stedy} implies that no nontrivial steady states 
with finite particle number
exist in the bosonic Yang–Gaudin model in the triplet sector or in the fermionic  singlet sector. This contrasts with the bosonic singlet sector. In the two-body case, a similar analysis can also be carried out by solving the Schrödinger equation in center-of-mass and relative coordinates (see Appendix C).

\subsection{Triplet solutions of the bosonic Yang-Gaudin model}
\label{subsection:3_3}
We consider the two-boson system in the triplet sector. We show that if the right eigenvalue of $H_{\mathrm{eff}}$ is real, then $c'$ must be real. From Eqs.~\eqref{eq_bethe_boson_1} and \eqref{eq_bethe_boson_2}, the Bethe equations reduce to those of the two-particle Lieb-Liniger model~\cite{re:LiebLiniger}
\begin{equation}\label{eq:2body_bethe_ParaBoson_1}
     e^{ik_1L}
    =\frac{k_1-k_2+ic'}{k_1-k_2-ic'},
\end{equation}
\begin{equation}\label{eq:2body_bethe_ParaBoson_2}
     e^{ik_2L}
    =\frac{k_2-k_1+ic'}{k_2-k_1-ic'}.
\end{equation}
From Eqs.~\eqref{eq:2body_bethe_ParaBoson_1} and \eqref{eq:2body_bethe_ParaBoson_2}, we obtain
\begin{equation}
\label{eq:2body_ParaBoson_1}
\begin{split}
    &e^{i(k_1+k_2)L}
    =\frac{k_1-k_2+ic'}{k_1-k_2-ic'}
    \frac{k_2-k_1+ic'}{k_2-k_1-ic'}
    =1\\
    &\Leftrightarrow
    \exists n\in\mathbb{Z},
    k_1+k_2=\frac{2n\pi}{L}.
\end{split}
\end{equation}
We now consider the condition on $c'$ under which the right eigenvalue of $H_{\mathrm{eff}}$
\begin{equation}
    E=\frac{\hbar^2}{2m}(k_1^2+k_2^2)
\end{equation}
is real. Writing $k_1=x_1+iy_1$ and $k_2=x_2+iy_2;(x_1,$ $y_1,x_2,y_2\in\mathbb{R})$, Eq.~\eqref{eq:2body_ParaBoson_1} implies
\begin{equation}
    x_1+x_2+i(y_1+y_2)=\frac{2n\pi}{L}\in\mathbb{R},
\end{equation}
and hence $y_2=-y_1$. In this case,
\begin{equation}
    \frac{2m}{\hbar^2}E
    =x_1^2+x_2^2-2y_1^2+2(x_1-x_2)y_1i.
\end{equation}
Thus, if $E$ is real, either $x_1=x_2$ or $y_1=0$ must hold.
First, consider the case $x_1=x_2$. In this case, $k_2=k_1^*$. From Eq.~\eqref{eq:2body_bethe_ParaBoson_1}, we have
\begin{equation}
    e^{ik_1L}
    =\frac{k_1-k_1^*+ic'}{k_1-k_1^*-ic'}
    =\frac{2y_1+c'}{2y_1-c'}.
\end{equation}
Taking the complex conjugate of both sides yields
\begin{equation}
    e^{-ik_1^*L}
    =\frac{2y_1+c'^*}{2y_1-c'^*},
\end{equation}
\begin{equation}
    \therefore e^{ik_1^*L}
    =\frac{2y_1-c'^*}{2y_1+c'^*}.  
\end{equation}
On the other hand, from Eq.~\eqref{eq:2body_bethe_ParaBoson_2}, we have
\begin{equation}
     e^{ik_1^*L}
    =\frac{k_1^*-k_1+ic'}{k_1^*-k_1-ic'}
    =\frac{2y_1-c'}{2y_1+c'},
\end{equation}
and therefore
\begin{equation}
    \frac{2y_1-c'^*}{2y_1+c'^*}
    =\frac{2y_1-c'}{2y_1+c'},
\end{equation}
\begin{equation}
    \therefore y_1(c'-c'^*)=0.
\end{equation}
If $y_1=0$, then $y_2=-y_1=0$, which implies $k_1=k_2$, contradicting the assumption that $k_1$ and $k_2$ are distinct. Hence $y_1\neq0$, and therefore $c'=c'^*$, and $c'$ is real.
Next, consider the case $y_1=0$. In this case, $y_2=0$, and thus $k_1,k_2\in\mathbb{R}$. From Eq.~\eqref{eq:2body_bethe_ParaBoson_1}, we obtain
\begin{equation}
    \frac{k_1-k_2-ic'}{k_1-k_2+ic'}
    =e^{-ik_1L}
    =\frac{k_1-k_2-ic'^*}{k_1-k_2+ic'^*},
\end{equation}
\begin{equation}
    \therefore (k_1-k_2)(c'-c'^*)=0.
\end{equation}
Since $k_1\neq k_2$, it follows that $c'=c'^*$, and hence $c'$ is real. Therefore,
\begin{equation}
    E\in\mathbb{R}
    \Rightarrow
    c'\in\mathbb{R}.
\end{equation}
Since $c'=2m(c-i\hbar\gamma/4)/\hbar^2$, this implies that the right 
eigenvalue is necessarily complex when $\gamma>0$.

\subsection{Singlet solutions of the fermionic Yang-Gaudin model}
\label{subsection:3_4}
We consider the two-body fermionic system in the singlet case. We show that if the right eigenvalue of $H_{\mathrm{eff}}$ is real, then $c'$ must also be real. From Eqs.~\eqref{eq_bethe_fermion_1} and \eqref{eq_bethe_fermion_2}, the Bethe equations are given by
\begin{equation}\label{eq:2body_bethe_AntiFermi_1}
    e^{ik_1L}
    =\frac{k_1-l+ic'/2}{k_1-l-ic'/2},
\end{equation}
\begin{equation}\label{eq:2body_bethe_AntiFermi_2}
    e^{ik_2L}
    =\frac{k_2-l+ic'/2}{k_2-l-ic'/2},
\end{equation}
\begin{equation}\label{eq:2body_bethe_AntiFermi_3}
    \frac{l-k_1+ic'/2}{l-k_1-ic'/2}
    \frac{l-k_2+ic'/2}{l-k_2-ic'/2}
    =1.
\end{equation}
From Eqs.~\eqref{eq:2body_bethe_AntiFermi_1}, \eqref{eq:2body_bethe_AntiFermi_2}, and \eqref{eq:2body_bethe_AntiFermi_3}, we obtain
\begin{equation}
\label{eq:2body_AntiFermi_1}
\begin{split}
    &e^{i(k_1+k_2)L}=1\\
    &\Leftrightarrow
    \exists n\in\mathbb{Z},
    k_1+k_2=\frac{2n\pi}{L}.
\end{split}
\end{equation}
Substituting Eq.~\eqref{eq:2body_AntiFermi_1} into Eq.~\eqref{eq:2body_bethe_AntiFermi_3}, we obtain
\begin{equation}
\begin{split}
    &\frac{l-k_1+ic'/2}{l-k_1-ic'/2}
    \frac{l+k_1-\frac{2n\pi}{L}+ic'/2}{l+k_1-\frac{2n\pi}{L}-ic'/2}
    =1\\
    &\Rightarrow
    l=\frac{n\pi}{L}
\end{split}
\end{equation}
We now consider the condition on $c'$ under which the right eigenvalue of $H_{\mathrm{eff}}$,
\begin{equation}
    E=\frac{\hbar^2}{2m}(k_1^2+k_2^2)
\end{equation}
is real. Writing $k_1=x_1+iy_1$ and $k_2=x_2+iy_2\;(x_1,y_1,$ $x_2,y_2\in\mathbb{R})$, Eq.~\eqref{eq:2body_AntiFermi_1} implies
\begin{equation}
    x_1+x_2+i(y_1+y_2)=\frac{2n\pi}{L}\in\mathbb{R},
\end{equation}
which leads to $y_2=-y_1$. Then
\begin{equation}
    \frac{2m}{\hbar^2}E
    =x_1^2+x_2^2-2y_1^2+2(x_1-x_2)y_1i.
\end{equation}
Hence, if $E$ is real, either $x_1=x_2$ or $y_1=0$ must hold. First, consider the case $x_1=x_2$. Then $k_2=k_1^*$. From Eq.~\eqref{eq:2body_bethe_AntiFermi_1}, we have
\begin{equation}
    e^{-ik_1^*L}
    =\frac{k_1^*-l-ic'^*/2}{k^*_1-l+ic'^*/2},
\end{equation}
\begin{equation}
    \therefore e^{ik_1^*L}
    =\frac{k_1^*-l+ic'^*/2}{k^*_1-l-ic'^*/2}.
\end{equation}
On the other hand, from Eq.~\eqref{eq:2body_bethe_AntiFermi_2},
\begin{equation}
    e^{ik_1^*L}
    =\frac{k_1^*-l+ic'/2}{k^*_1-l-ic'/2}.
\end{equation}
Thus,
\begin{equation}
    \frac{k_1^*-l+ic'^*/2}{k^*_1-l-ic'^*/2}
    =\frac{k_1^*-l+ic'/2}{k^*_1-l-ic'/2},
\end{equation}
which leads to
\begin{equation}
    (k_1^*-l)(c'-c'^*)=0.
\end{equation}
Assuming $c'\neq c'^*$, we obtain $k_1^*=l\in\mathbb{R}$, which implies $k_1^*=k_1$. Together with $k_2=k_1^*$, this leads to $k_1=k_2$, contradicting the assumption that $k_1$ and $k_2$ are distinct. Therefore, $c'=c'^*$, and $c'\in\mathbb{R}$.
Next, consider the case $y_1=0$. Then $y_2=-y_1=0$, and thus $k_1,k_2\in\mathbb{R}$. From Eq.~\eqref{eq:2body_bethe_AntiFermi_1},
\begin{equation}
    e^{-ik_1L}
    =\frac{k_1-l-ic'/2}{k_1-l+ic'/2}.
\end{equation}
Taking the complex conjugate, we also have
\begin{equation}
    e^{-ik_1L}
    =\frac{k_1-l-ic'^*/2}{k_1-l+ic'^*/2}.
\end{equation}
Hence,
\begin{equation}
    \frac{k_1-l-ic'/2}{k_1-l+ic'/2}
    =\frac{k_1-l-ic'^*/2}{k_1-l+ic'^*/2},
\end{equation}
which gives
\begin{equation}
    (k_1-l)(c'-c'^*)=0.
\end{equation}
Applying the same argument to Eq.~\eqref{eq:2body_bethe_AntiFermi_2}, we obtain
\begin{equation}
    (k_2-l)(c'-c'^*)=0.
\end{equation}
If $c'\neq c'^*$, then $k_1=l=k_2$, which again contradicts the assumption that $k_1\neq k_2$. Therefore, $c'=c'^*$, and hence $c'\in\mathbb{R}$. Thus, we conclude that
\begin{equation}
    E\in\mathbb{R}
    \Rightarrow
    c'\in\mathbb{R}.
\end{equation}
As in the bosonic triplet case, this implies that the right eigenvalue becomes
complex in the presence of dissipation.

\subsection{String solutions of the fermionic Yang-Gaudin model}
In isolated integrable systems with attractive interactions $c<0$ and large system size $L$,  solutions known as string solutions can emerge~\cite{re:one_dimensional_hubbard}. 
Here we discuss the existence of string solutions in the two-body fermionic Yang–Gaudin model in the presence of dissipation.

\begin{figure}
  \centering
  \includegraphics[width=0.45\textwidth]{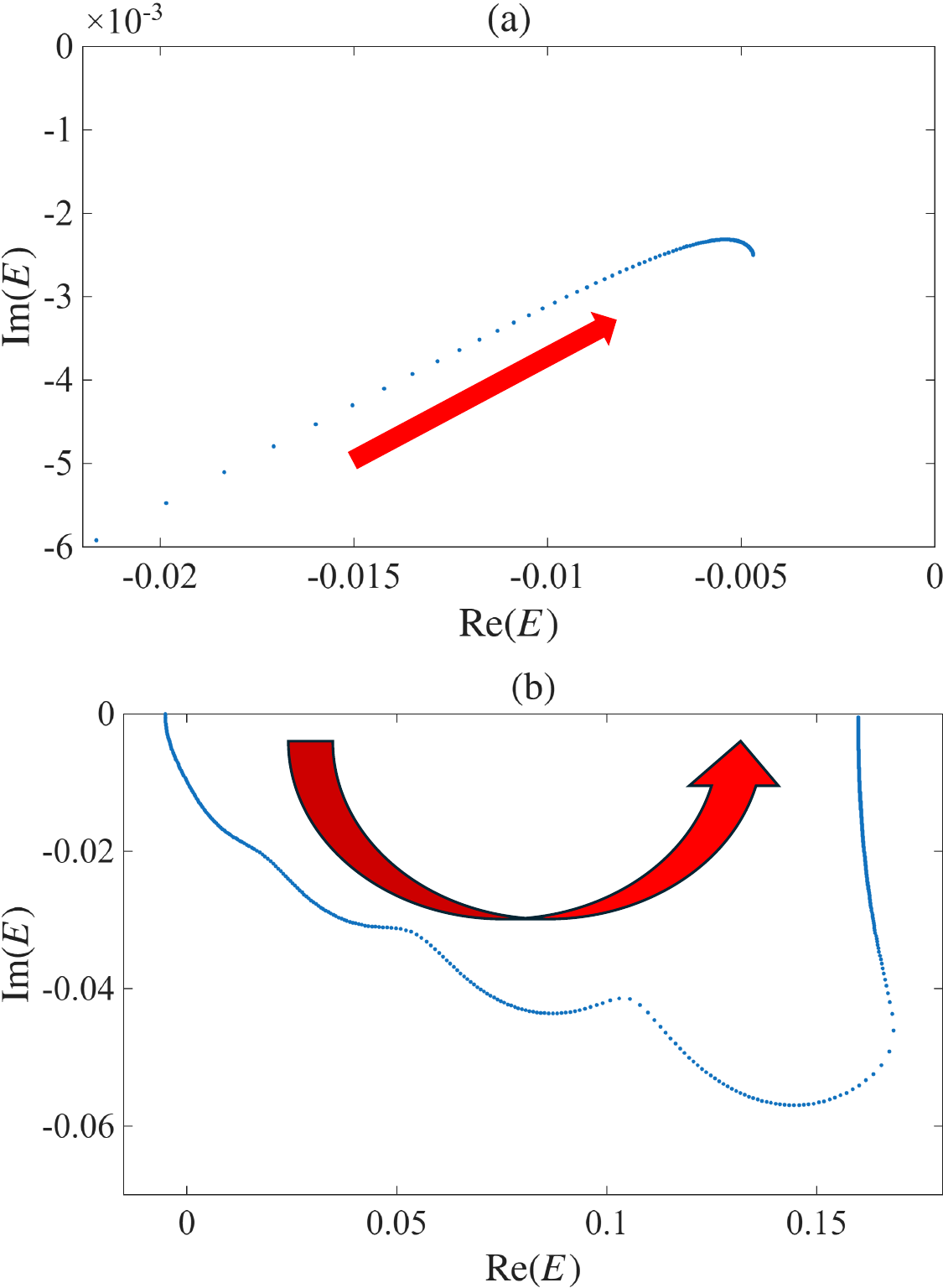}
  \caption{Numerical solutions of right eigenvalue, for $2m=\hbar=1$. (a) Dots shows right eigenvalues for $c=-0.1$ and $\gamma=0.1$, as $L$ is increased from $10$ to $300$, the right eigenvalues evolve in the direction indicated by the red arrows. In the limit of large $L$, $E$ converges to $-c'^2/2$.
(b) Dots shows right eigenvalues for $c=-0.1$ and $L=100$, as $\gamma$ is increased from $0$ to $100$, the right eigenvalues evolve in the direction indicated by the red arrows.}
  \label{fig:string}
\end{figure}

The Bethe equation Eq.~\eqref{eq:2body_bethe_AntiFermi_1} can be written as
\begin{equation}
    e^{-\mathrm{Im}(k_1)L}e^{i\mathrm{Re}(k_1)L}
    =\frac{k_1-l+ic'/2}{k_1-l-ic'/2}.
\end{equation}
Assuming that $k_1$ remains finite,  the left-hand side tends to $0$ as $L\rightarrow\infty$
when $\mathrm{Im}(k_1)>0$, whereas it diverges when $\mathrm{Im}(k_1)<0$. Therefore,
\begin{equation}
    k_1=
    \begin{cases}
        l-ic'/2+O(e^{-\mathrm{Im}(k_1)L}),
        &(\mathrm{Im}(k_1)>0)\\
        l+ic'/2+O(e^{-|\mathrm{Im}(k_1)|L}),
        &(\mathrm{Im}(k_1)<0).
    \end{cases}
\end{equation}
From Eq.~\eqref{eq:2body_AntiFermi_1}, we consider solutions satisfying $k_1+k_2=0$. In this case, Eq.~\eqref{eq:2body_bethe_AntiFermi_3} implies $l=0$, and the corresponding right eigenvalue is given by
\begin{equation}
    E=-\frac{\hbar^2}{2m}\frac{c'^2}{2}
    +O(e^{-|\mathrm{Im}(k_1)|L})
\end{equation}
This shows that for sufficiently large $L$ and small $\gamma$, $\mathrm{Re}(E)<0$, indicating a bound state.

We have numerically confirmed the existence of such solutions in the singlet sector of the two-body fermionic Yang–Gaudin model. Figure~\ref{fig:string}(a) illustrates  a string solution for which $E\rightarrow -c'^2/2$ as $L$ becomes large. Furthermore, Fig.~\ref{fig:string}(b) shows that, for the string solution,  $\mathrm{Re}(E)$ becomes positive as $\gamma$ increases, while $\mathrm{Im}(E)\rightarrow 0$ in the limit $\gamma\rightarrow\infty$. This behavior can be interpreted as a manifestation of the quantum Zeno effect.

\section{THE MANY-BODY PROBLEM}

\begin{figure*}[t]
  \centering
  \includegraphics[width=1
  \textwidth]{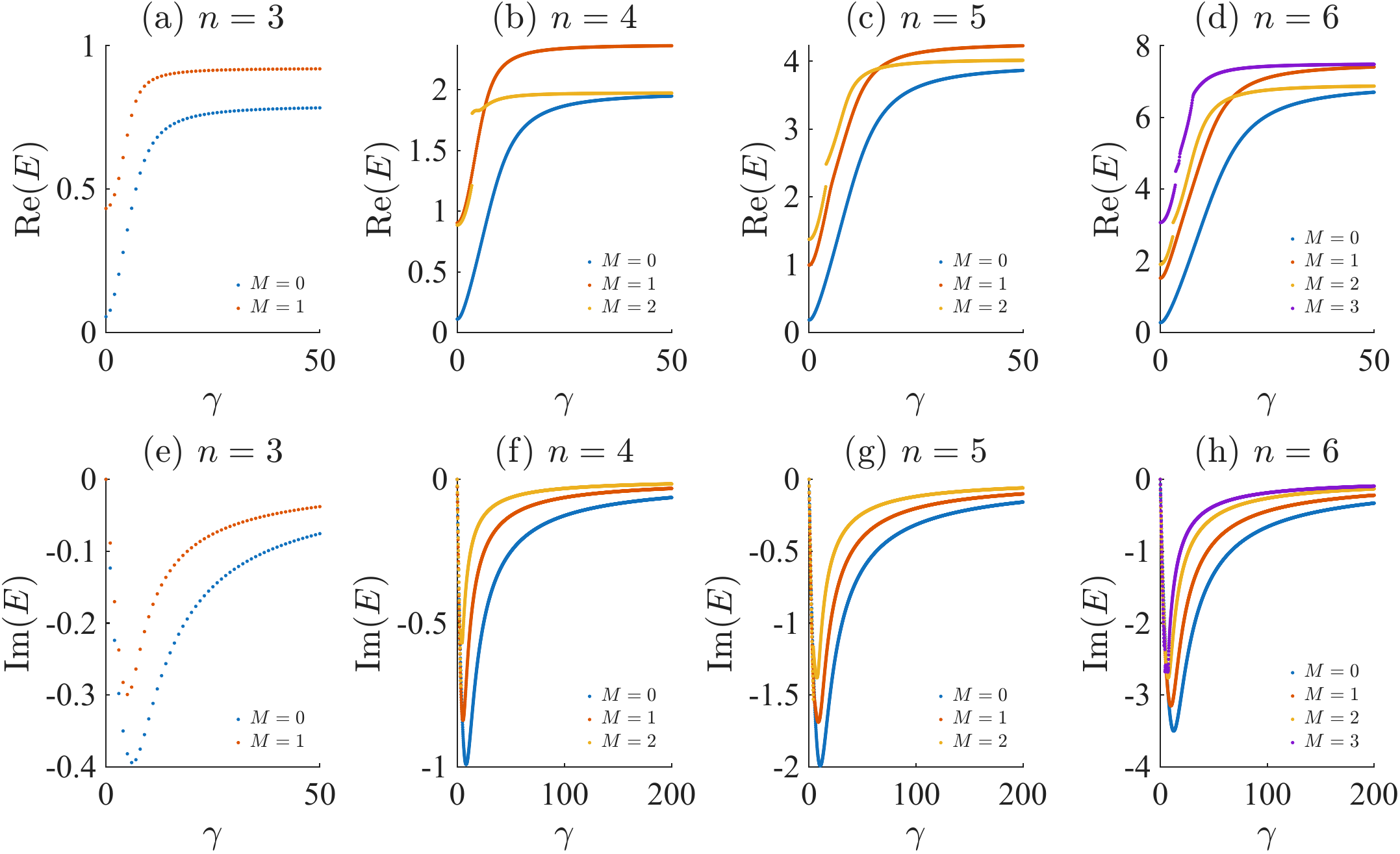}
  \caption{Numerical results for the right eigenvalues of $H_{\mathrm{eff}}$ in the bosonic Yang–Gaudin model, for $c=1, L=10$. Panels (a)-(d) show the real parts of the right eigenvalues as functions of $\gamma$ for $n=3,4,5,6$, respectively. 
  Panels(e)-(h) show the corresponding imaginary parts of the right eigenvalues as functions of $\gamma$ for $n=3,4,5,6$, respectively.}
  \label{fig:boson}
\end{figure*}

\begin{figure*}[t]
  \centering
  \includegraphics[width=1\textwidth]{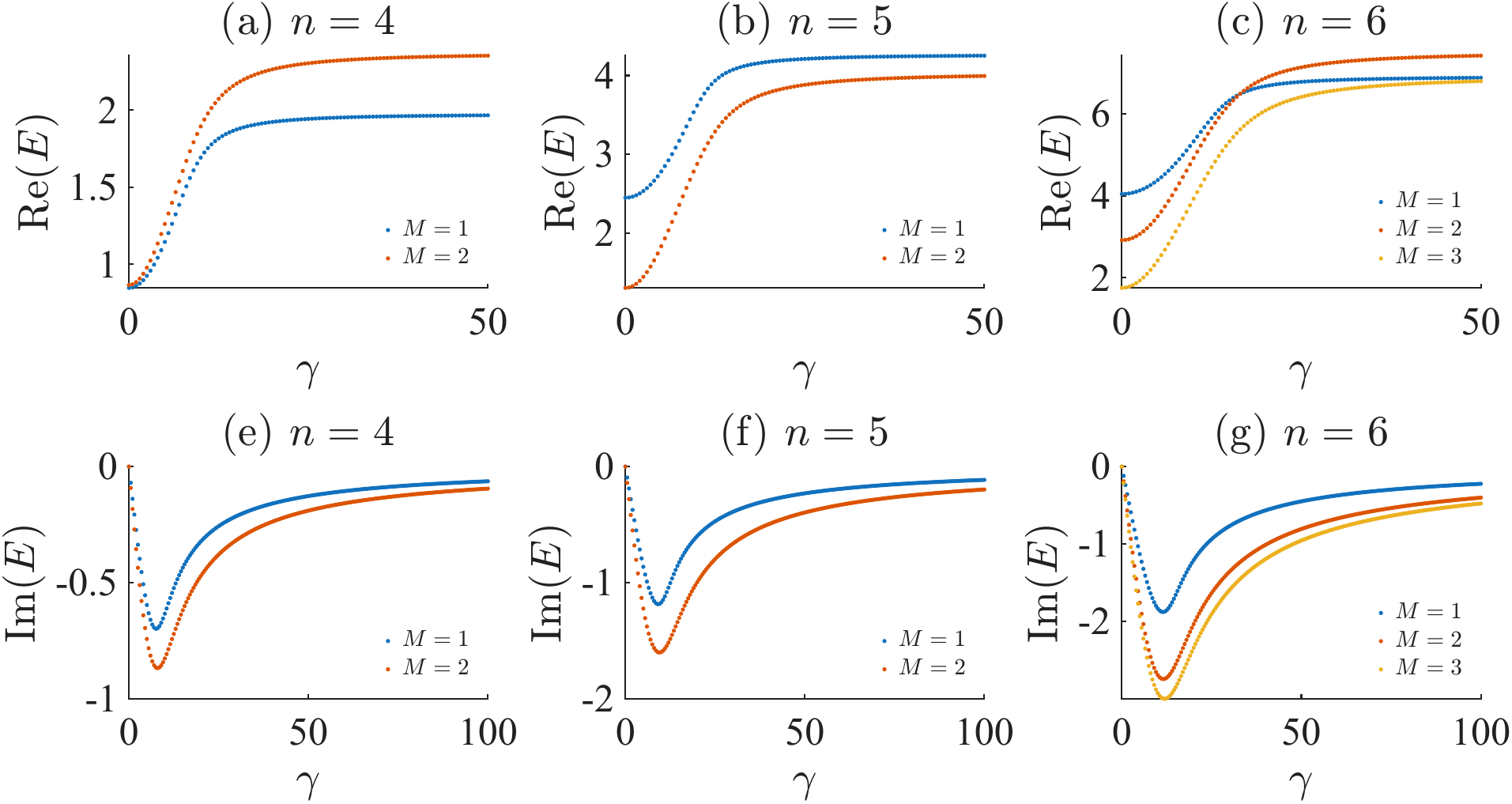}
  \caption{Numerical results for the right eigenvalues of $H_{\mathrm{eff}}$ in the fermionic Yang–Gaudin model, for $c=1, L=10$. Panels (a)-(c) show the real parts of the right eigenvalues as functions of $\gamma$ for $n=4,5,6$, respectively. Panels (e)-(g) show the imaginary parts of the right eigenvalues as functions of $\gamma$ for $n=4,5,6$, respectively.}
  \label{fig:fermion}
\end{figure*}

In this section, we discuss the differences between bosonic and fermionic systems with three or more particles. To this end, we numerically solve the Bethe equations Eqs.~\eqref{eq:Betheeq_qu1}, \eqref{eq:Betheeq_qu2}, \eqref{eq:Betheeq_qu3}, and \eqref{eq:Betheeq_qu4}. 
The sets of quantum numbers $I_j$ and $J_{\alpha}$ are chosen to be mutually distinct and as close to zero as possible.
For example, in the bosonic case with $n=6$ and $M=2$, we take $I_j=\lbrace-5/2,-3/2,-1/2,1/2,3/2,5/2\rbrace$ and $J_\alpha=\lbrace-1/2,1/2\rbrace$. Similarly, for $n=6$ and $M=3$, we take $I_j=\lbrace-2,-1,0,1,2,3\rbrace$ and $J_\alpha=\lbrace-1,0,1\rbrace$. 
Throughout this section, We set $2m=\hbar=1$.

Figure~\ref{fig:boson} shows the dependence of the real and imaginary parts of the right eigenvalues on $\gamma$ in the bosonic case.
First, in the absence of dissipation, except for the cases $n=4$ with $M=1,2$, the energy increases as $M$ increases within the range $M\leq n/2$. Even for $n=4$, the values for $M=1$ and $M=2$ are close to each other and both are larger than that for $M=0$. This indicates that, in the absence of dissipation, bosonic states with smaller $M$ are more stable in the sense that they have lower energy. The fact that for $n=4$ the $M=2$ state has lower energy than the $M=1$ state may be attributed to its higher symmetry.
By contrast, in the presence of dissipation, the imaginary part of the right eigenvalues becomes less negative as $M$ increases. 
Equivalently, the magnitude of the particle-loss rate decreases with increasing $M$.
This implies that states with larger $M$ are more stable in the dissipative case.

Figure~\ref{fig:fermion} shows the dependence of the real and imaginary parts of the right eigenvalues on $\gamma$ in the fermionic case. In the absence of dissipation, for $n=4$, the energy at $M=1$ is slightly higher than that at $M=2$. For $n=5$ and $6$, the energy decreases as $M$ increases within the range $M\leq n/2$, indicating that larger $M$ corresponds to more stable states. By contrast, in the presence of dissipation, the imaginary part of the right eigenvalues becomes less negative as $M$ decreases.
Equivalently, the magnitude of the particle-loss rate decreases as $M$ decreases.
Hence, states with smaller $M$ are more stable in the dissipative case.
In other words, dissipation favors antiferromagnetic-like configurations in bosonic systems
and ferromagnetic-like configurations in fermionic systems.

These results show that, for both bosons and fermions, the ordering of stability with respect to $M$ is reversed by dissipation. 
In the  dissipative case, bosonic systems become more stable as $M$ increases (within $M\leq n/2$), whereas fermionic systems become more stable as $M$ decreases.

A feature common to both bosons and fermions is that the imaginary part of the right eigenvalues tends to become more negative as the particle number $n$ increases. Furthermore, for sufficiently large $\gamma$, the imaginary part of the right eigenvalues approaches zero. This behavior can be interpreted as a manifestation of the quantum Zeno effect.

\section{CONCLUSION}
We have shown that the one-dimensional dissipative Yang–Gaudin model with two-body loss is exactly solvable. In particular, as a generalization of \cite{re:Torres}, we have rigorously established the relationship between the spectrum of the Liouvillian superoperator in an infinite-dimensional Hilbert space and that of the non-Hermitian Hamiltonian $H_{\mathrm{eff}}$. Furthermore, as a generalization of \cite{re:dis_Tonks_the}, we have derived the relation between the initial particle-loss rate  and the right eigenvalues of $H_{\mathrm{eff}}$.
For the two-body bosonic system in the singlet sector, we proved that the right eigenvalues of $H_{\mathrm{eff}}$ are real even in the presence of dissipation, implying the existence of steady-state solutions of the master equation. In contrast, for the bosonic triplet case and the fermionic singlet case, we showed that dissipation renders the right eigenvalues of $H_{\mathrm{eff}}$ complex.
As a many-body effect, we demonstrated that dissipation changes the  stable spin configuration: in bosonic systems, the stable configuration shifts from ferromagnetic to antiferromagnetic, whereas in fermionic systems it shifts from antiferromagnetic to ferromagnetic. 
Equation~\eqref{eq:relarion_loss_E_2} further shows that, for a pure initial state,
the initial particle-loss rate directly probes the imaginary part of the corresponding
right eigenvalue of  $H_{\mathrm{eff}}$, highlighting  the physical significance of the effective non-Hermitian 
description.
These results show that dissipation in the Yang-Gaudin model does not merely induce decay, but can qualitatively
reorganize spin stability while preserving exact solvability.

\section*{acknowledgments}
We are grateful to M. Nakagawa for discussions.
S.U. is supported by JST PRESTO~(JPMJPR2351) and
JSPS KAKENHI~(JP25K07191).

\appendix

    \section{EXACT LIOUVILLIAN SPECTRUM}
In this Appendix, we extend the method of Ref.~\cite{re:Torres} to an infinite-dimensional Hilbert space and derive Eqs.~\eqref{eigenvalue_K} and \eqref{inclusion}. We impose the following assumptions:
 (a)  $\varsigma(\mathcal{K})=\varsigma_P(\mathcal{K})$;
 (b)  $\mathcal{A}$ is bounded. Let $\mathcal{H}^*$ be the dual space of $\mathcal{H}$. Then $\mathcal{K}$ is a linear operator acting on $\mathcal{H}\otimes\mathcal{H}^*$. From Eq.~\eqref{eq:comp_1}, for an arbitrary linear operator $A$ on $\mathcal{H}\otimes\mathcal{H}^*$ we have
\begin{equation}
    A=\sum_{n,n',j,j'}\Braket{q_j^n|A|q_{j'}^{n'}}
    \Ket{r_j^n}\Bra{r_{j'}^{n'}},
\end{equation}
and therefore $\lbrace\ket{r_j^n}\bra{r_{j'}^{n'}}\rbrace$ forms a complete set in $\mathcal{H}\otimes\mathcal{H}^*$. We define
\begin{equation}
    \Ket{\Ket{R_{j,k}^{n,l}}\!}
    =\ket{r_j^n}\bra{r_{k}^{l}},
\end{equation}
and define a linear functional $\Bra{\!\Bra{Q_{j,k}^{l,n}}}:\mathcal{H}\otimes\mathcal{H}^*\rightarrow\mathbb{C}$ by
\begin{equation}
    \Bra{\!\Bra{Q_{j,k}^{n,l}}}A
    =\mathrm{tr}\left[
    \left(\Ket{q_j^n}\Bra{q_{k}^{l}}\right)^\dagger A
    \right],
\end{equation}
for $A\in\mathcal{H}\otimes\mathcal{H}^*$. Then
\begin{equation}
\begin{split}
    &\Braket{\!\Braket{Q_{j,k}^{n,l}|R_{j',k'}^{n',l'}}\!}
    =\mathrm{tr}\left[
    \left(\Ket{q_j^n}\Bra{q_{k}^{l}}\right)^\dagger 
    \ket{r_{j'}^{n'}}\bra{r_{k'}^{l'}}\right]\\
    &=\Braket{q_j^n|r_{j'}^{n'}}\Braket{q_{k}^{l}|r_{k'}^{l'}}
    =\delta_{n,n'}\delta_{l,l'}\delta_{j,j'}\delta_{k,k'}.
\end{split}
\end{equation}
Since this and $\lbrace\ket{r_j^n}\bra{r_{j'}^{n'}}\rbrace$ form a complete set of $\mathcal{H}\otimes\mathcal{H}^*$, we obtain
\begin{equation}
\label{eq:comp_2}
    \sum_{n,l,j,k}
    \Ket{\Ket{R_{j,k}^{n,l}}\!}\Bra{\!\Bra{Q_{j,k}^{n,l}}}
    =\mathbf{I}_{\mathcal{H}\otimes\mathcal{H}^*}.
\end{equation}
Therefore,
\begin{equation}
    \mathcal{K}
    =\sum_{n,l,j,k}\frac{1}{i\hbar}\left(
    \varepsilon_{j}^{(n)}-\varepsilon_{k}^{(l)*}
    \right)
    \Ket{\Ket{R_{j,k}^{n,l}}\!}\Bra{\!\Bra{Q_{j,k}^{n,l}}},
\end{equation}
and thus we obtain
\begin{equation}
    \varsigma_P(\mathcal{K})
    =\left\lbrace
    \frac{1}{i\hbar}\left(
    \varepsilon_{j}^{(n)}-\varepsilon_{k}^{(l)*}
    \right)
    \right\rbrace_{\substack{n,l=0,1,2,\cdots,N\\j,k=1,2,\cdots}}.
\end{equation}
From Eq.~\eqref{eq:comp_2}, we obtain
\begin{equation}
    \mathcal{A}
    =\sum_{n,l,j,k}\sum_{n',l',j',k'}
    \Braket{\!\Braket{Q_{j,k}^{n,l}|\mathcal{A}|R_{j',k'}^{n',l'}}\!}
    \Ket{\Ket{R_{j,k}^{n,l}}\!}\Bra{\!\Bra{Q_{j',k'}^{n',l'}}}.
\end{equation}
Here we write the action of $\mathcal{A}$ more explicitly in terms of the wave functions. From Eq.~\eqref{eq:def_A}, we obtain
\begin{equation}
\begin{split}
    &\mathcal{A}\Ket{\Ket{R_{j',k'}^{n',l'}}\!}\\
    &=\gamma\sum_{\sigma_\alpha,\sigma_\beta}
    \int dx\;\hat{\psi}_{\sigma_\alpha}(x)
    \hat{\psi}_{\sigma_\beta}(x)
    \ket{r_{j'}^{n'}}\bra{r_{k'}^{l'}}
    \hat{\psi}^{\dagger}_{\sigma_\beta}(x)
    \hat{\psi}^{\dagger}_{\sigma_\alpha}(x).
\end{split}
\end{equation}
Let $\phi_{r_{j}^n}(x_1,\cdots,x_n,\sigma_1,\cdots,\sigma_n)$ denote the amplitude of $\ket{\sigma_1,\cdots,\sigma_n}$ in the wave function corresponding to the right eigenstate of $H_{\mathrm{eff}}$ with particle number $n$. The right eigenket can be expressed in terms of the wave function as
\begin{widetext}
\begin{equation}
\begin{split}
    \ket{r_{j}^{n}}
    =\sum_{\sigma_1,\cdots,\sigma_n}
    \frac{1}{\sqrt{n!}}
    \int\!\! dx_1\cdots dx_{n} 
    \phi_{r_{j}^{n}}
    \hat{\psi}_{\sigma_1}^\dagger(x_1)
    \cdots\hat{\psi}_{\sigma_n}^\dagger(x_{n})\ket{0}.
\end{split}
\end{equation}
Then
\begin{equation}
\begin{split}
    &\hat{\psi}_{\sigma_a}(x)
    \hat{\psi}_{\sigma_b}(x)
    \ket{r_{j'}^{n'}}
    =\frac{1}{\sqrt{n'!}}
    \sum_{\sigma_1,\cdots,\sigma_{n'}}
    \int dx_1\cdots dx_{n'} \phi_{r_{j'}^{n'}}(x_1,\cdots,x_{n'},\sigma_1,\cdots,\sigma_{n'})
    \hat{\psi}_{\sigma_a}(x)
    \hat{\psi}_{\sigma_b}(x)
    \hat{\psi}^\dagger_{\sigma_1}(x_1)\cdots
    \hat{\psi}^\dagger_{\sigma_{n'}}(x_{n'})\ket{0}\\
    &=\frac{1}{\sqrt{n'!}}\sum_{\sigma_1,\cdots,\sigma_{n'}}
    \int dx_1\cdots dx_{n'} \phi_{r_{j'}^{n'}}(x_1,\cdots,x_{n'},\sigma_1,\cdots,\sigma_{n'})
    \hat{\psi}_{\sigma_a}(x)
    \sum_{\alpha=1}^{n'}(S_{pm})^{\alpha-1}
    \delta_{\sigma_b\sigma_\alpha}\delta(x-x_\alpha)
    \left(\prod_{\substack{\xi=1\\\xi\neq\alpha}}^{n'}
    \hat{\psi}_{\sigma_\xi}^\dagger(x_\xi)\right)\ket{0}\\
    &=\frac{1}{\sqrt{n'!}}\sum_{\alpha=1}^{n'}
    \sum_{\sigma_1,\cdots,\sigma_{n'}}
    \int dx_1\cdots dx_{n'} \phi_{r_{j'}^{n'}}(x_\alpha,x_1,\cdots,x_{\alpha-1},x_{\alpha+1},\cdots,x_{n'},
    \sigma_\alpha,\sigma_1,\cdots,\sigma_{\alpha-1},\sigma_{\alpha+1},\cdots,\sigma_{n'})\\
    &\;\;\;\;\;\;\;\;\;\;\;\;\;\;\;\;\;\;
    \;\;\;\;\;\;\;\;\;\;\;\;\;\;\;\times
    \hat{\psi}_{\sigma_a}(x)
    \delta_{\sigma_b\sigma_\alpha}\delta(x-x_\alpha)
    \left(\prod_{\substack{\xi=1\\\xi\neq\alpha}}^{n'}
    \hat{\psi}_{\sigma_\xi}^\dagger(x_\xi)\right)\ket{0}\\
    &=\frac{\sqrt{n'}}{\sqrt{(n'-1)!}}
    \sum_{\sigma_1,\cdots,\sigma_{n'-1}}
    \int dx_1\cdots dx_{n'-1} \phi_{r_{j'}^{n'}}(x,x_1,\cdots,x_{n'-1},\sigma_b,
    \sigma_1,\cdots,\sigma_{n'-1})
    \hat{\psi}_{\sigma_a}(x)
    \left(\prod_{\xi=1}^{n'-1}
    \hat{\psi}_{\sigma_\xi}^\dagger(x_\xi)\right)\ket{0}\\
    &=\frac{\sqrt{n'(n'-1)}}{\sqrt{(n'-2)!}}
    \sum_{\sigma_1,\cdots,\sigma_{n'-2}}
    \int dx_1\cdots dx_{n'-2} \phi_{r_{j'}^{n'}}(x,x,x_1,\cdots,x_{n'-2},\sigma_b,\sigma_a
    \sigma_1,\cdots,\sigma_{n'-2})
    \left(\prod_{\xi=1}^{n'-2}
    \hat{\psi}^\dagger(x_\xi)\right)\ket{0}\\
\end{split}
\end{equation}
where $S_{pm}$ takes the value $1$ for bosons and $-1$ for fermions.
If $n'<2$, the vector $\hat{\psi}^2(x)\ket{r_{j'}^{n'}}$ becomes the zero vector.
Using this result, we obtain
\begin{equation}
\label{eq:A_wave}
\begin{split}
    &\mathcal{A}\Ket{\Ket{R_{j',k'}^{n',l'}}\!}\\
    &=\gamma
    \sum_{\substack{\sigma_1,\cdots,\sigma_{n'-2}\\
    \sigma'_1,\cdots,\sigma'_{l'-2}}}
    \int dx_1\cdots dx_{n'-2} dx'_1\cdots dx'_{l'-2}\;
    \Psi_{r_{j'}^{n'},r_{k'}^{l'}}
    \hat{\psi}_{\sigma_1}^\dagger(x_1)\cdots\hat{\psi}_{\sigma_{n'-2}}^\dagger(x_{n'-2})\ket{0}
    \bra{0}\hat{\psi}_{\sigma'_{l'-2}}(x'_{l'-2})\cdots
    \hat{\psi}_{\sigma'_{1}}(x'_1),
\end{split}
\end{equation}
where
\begin{equation}
\begin{split}
    &\Psi_{r_{j'}^{n'},r_{k'}^{l'}}(x_1,\cdots,x_{n'-2},x'_1,\cdots,x'_{l'-2},
    \sigma_1,\cdots,\sigma_{n'-2},
    \sigma'_1,\cdots,\sigma'_{l'-2})\\
    &=\frac{\sqrt{n'(n'-1)l'(l'-1)}}{\sqrt{(n'-2)!(l'-2)!}}
    \sum_{\sigma_a,\sigma_b}
    \int dx\;\phi_{r_{j'}^{n'}}(x,x,x_1,\cdots,x_{n'-2},\sigma_b,\sigma_a
    \sigma_1,\cdots,\sigma_{n'-2})\\
    &\;\;\;\;\;\;\;\;\;\;\;\;\;\;\;
    \;\;\;\;\;\;\;\;\;\;\;\;\;\;\;\;
    \;\;\;\;\;\;\;\;\;\;\;\;\;\;\times
    \phi^*_{r_{k'}^{l'}}(x,x,x'_1,\cdots,x'_{l'-2},\sigma_b,\sigma_a
    \sigma'_1,\cdots,\sigma'_{l'-2}).
\end{split}
\end{equation}
From this expression we find
\begin{equation}
    \mathcal{A}
    \Ket{\Ket{R_{j',k'}^{n',l'}}\!}
    \ket{r_{j''}^{n''}}
    =\gamma
    \sum_{\substack{\sigma_1,
    \cdots,\sigma_{n'-2}\\
    \sigma'_1,\cdots,\sigma'_{l'-2}}}
    \int dx_1\cdots dx_{n'-2} \Psi'_{r_{j'}^{n'},r_{k'}^{l'}}
    \hat{\psi}_{\sigma_1}^\dagger(x_1)\cdots
    \hat{\psi}_{\sigma_{n'-2}}^\dagger(x_{n'-2})\ket{0},
\end{equation}
where
\begin{equation}
    \Psi'_{r_{j'}^{n'},r_{k'}^{l'}}
    =\int dx'_1\cdots dx'_{l'-2}\;
    \Psi_{r_{j'}^{n'},r_{k'}^{l'}}
    \Braket{0|\hat{\psi}
    _{\sigma'_{l'-2}}(x'_{l'-2})\cdots\hat{\psi}
    _{\sigma'_1}(x'_1)|r_{j''}^{n''}}\delta_{n'',l'-2},
\end{equation}
which implies
\begin{equation}
    \mathcal{A}\Ket{\Ket{R_{j',k'}^{n',l'}}\!}
    \in \mathcal{H}_{n'-2}\otimes\mathcal{H}_{l'-2}^*.
\end{equation}
Therefore,
\begin{equation}
    \mathcal{A}\Ket{\Ket{R_{j',k'}^{n',l'}}\!}
    =\sum_{j'',k''}\mathcal{A}_{j'',k''}^{(n'-2,l'-2)}
    \Ket{\Ket{R_{j'',k''}^{n'-2,l'-2}}\!},
\end{equation}
and hence
\begin{equation}
\begin{split}
    \mathcal{A}
    &=\sum_{n,l,j,k}\sum_{n',l',j',k'}
    \sum_{j'',k''}\mathcal{A}_{j'',k''}^{(n'-2,l'-2)}
    \delta_{n,n'-2}\delta_{l,l'-2}\delta_{j,j''}\delta_{k,k''}\Ket{\Ket{R_{j,k}^{n,l}}\!}\Bra{\!\Bra{Q_{j',k'}^{n',l'}}}\\
    &=\sum_{n',l',j,k,j',k'}
    \mathcal{A}_{j,k}^{(n'-2,l'-2)}
    \Ket{\Ket{R_{j,k}^{n'-2,l'-2}}\!}\Bra{\!\Bra{Q_{j',k'}^{n',l'}}}.
\end{split}
\end{equation}
\end{widetext}
Let $\rho_r(\mathcal{L})$ and $\rho_r(\mathcal{K})$ denote the resolvent sets of $\mathcal{L}$ and $\mathcal{K}$, respectively. If $\lambda_e\notin\varsigma(\mathcal{K})$, then $\lambda_e\in\rho_r(\mathcal{K})$ and hence $\mathcal{K}'\equiv(\lambda_e\mathbf{I}_{\mathcal{H}\otimes\mathcal{H}^*}-\mathcal{K})^{-1}$ exists,
\begin{equation}
    \mathcal{K}'
    =\sum_{n,l,j,k}
    \frac{1}{\lambda_e-\frac{1}{i\hbar}\left(
    \varepsilon_{j}^{(n)}-\varepsilon_{k}^{(l)*}
    \right)}
    \Ket{\Ket{R_{j,k}^{n,l}}\!}\Bra{\!\Bra{Q_{j,k}^{n,l}}}.
\end{equation}
Then
\begin{equation}
\begin{split}
    \lambda_e\mathbf{I}_{\mathcal{H}\otimes\mathcal{H}^*}-\mathcal{L}
        &=\lambda_e\mathbf{I}_{\mathcal{H}\otimes\mathcal{H}^*}-\mathcal{K}-\mathcal{A}\\
        &=(\lambda_e\mathbf{I}_{\mathcal{H}\otimes\mathcal{H}^*}-\mathcal{K})(\mathbf{I}_{\mathcal{H}\otimes\mathcal{H}^*}-\mathcal{K}'\mathcal{A}).
\end{split}
\end{equation}
Since $(\mathcal{K}'\mathcal{A})^N=O$,
\begin{equation}
    (\mathbf{I}_{\mathcal{H}\otimes\mathcal{H}^*}-\mathcal{K}'\mathcal{A})^{-1}
    =\sum_{j=0}^{N-1}(\mathcal{K}'\mathcal{A})^j,
\end{equation}
and therefore $(\lambda_e\mathbf{I}_{\mathcal{H}\otimes\mathcal{H}^*}-\mathcal{L})^{-1}$ exists as
\begin{equation}
    (\lambda_e\mathbf{I}_{\mathcal{H}\otimes\mathcal{H}^*}-\mathcal{L})^{-1}
    =\sum_{j=0}^{N-1}(\mathcal{K}'\mathcal{A})^j
    \mathcal{K}'.
\end{equation}
Since $\mathcal{A}$ is bounded, $(\lambda_e\mathbf{I}_{\mathcal{H}\otimes\mathcal{H}^*}-\mathcal{L})^{-1}$ is also bounded, which implies $\lambda_e\in\rho_r(\mathcal{L})$, i.e., $\lambda_e\notin\varsigma(\mathcal{L})$. Therefore,
\begin{equation}
    \varsigma_P(\mathcal{L})\subset
    \varsigma(\mathcal{L})\subset\varsigma(\mathcal{K})
    =\varsigma_P(\mathcal{K}).
\end{equation}

\section{ALGEBRAIC BETHE ANSATZ FOR THE YANG-GAUDIN MODEL}
In this Appendix, we show that the right eigenstates obtained by the Bethe ansatz are highest-weight states. The notation of the algebraic Bethe ansatz follows Refs.~\cite{re:ABA_FH,re:ABA_BYG}.

\subsection{Eigenstates of the bosonic Yang-Gaudin model}
We first derive eigenstates of the bosonic Yang-Gaudin model. From the continuity of the wavefunction Eq.~\eqref{eq:bethe_wavefunction} and
its first derivative at equal positions as $x_j=x_{j+1}$, we have
\begin{equation}
\label{eq_B:bc}
\begin{split}
    &A_{\lbrace\sigma_{Q}\rbrace}
    (\lbrace k_{P\sigma_{j(j+1)}}
    \rbrace)\\
    &=\frac{(k_{P_j}-k_{P_{j+1}})
    A_{\lbrace\sigma_{Q\sigma_{j(j+1)}}\rbrace}(\lbrace k_{P}
    \rbrace)-ic'A_{\lbrace\sigma_{Q}\rbrace}(\lbrace k_{P}
    \rbrace)}{k_{P_j}-k_{P_{j+1}}+ic'},
\end{split}
\end{equation}
where $A_{\lbrace\sigma_{Q}\rbrace}(\lbrace k_{P}\rbrace)\equiv A_{\sigma_{Q_1}\cdots\sigma_{Q_n}}(k_{P_1},\cdots,k_{P_n})$ and $\sigma_{\alpha,\beta}$ denotes the transposition that exchanges $\alpha$ and $\beta$. From periodic boundary condition
\begin{equation}
    \psi(x_1,\cdots,x_{Q_1}+L,\cdots,x_n)
    =\psi(x_1,\cdots,x_{Q_1},\cdots,x_n)
\end{equation}
where 
\begin{equation}
\begin{split}
    &\psi(x_1,\cdots,x_{Q_1}+L,\cdots,x_n,\sigma_1,\cdots,\sigma_n)\\
    &\equiv\sum_P 
    A_{\lbrace\sigma_{Q\tau}\rbrace}(\lbrace k_{P}\rbrace)
    \exp\left[i\sum_{j=1}^n k_{P_j}x_{Q\tau(j)}\right]
    \exp[ik_{P_n}L],\\
    &\tau=
    \begin{pmatrix}
        1 & 2 & \cdots & n-1 & n \\
        2 & 3 & \cdots & n & 1
    \end{pmatrix},
\end{split}
\end{equation}
we have
\begin{equation}
\label{eq_B:pbc}
    A_{\lbrace\sigma_{Q}\rbrace}(\lbrace k_{P}\rbrace)
    =A_{\lbrace\sigma_{Q\tau}\rbrace}(\lbrace k_{P\tau}\rbrace)
    \exp[ik_{P_1}L].
\end{equation}

We next introduce
\begin{equation}
    \Ket{k_{P_1},\cdots,k_{P_n}}
    \equiv\sum_{\lbrace\sigma_i\rbrace=\uparrow,\downarrow}
    A_{\lbrace\sigma_{Q}\rbrace}
    (\lbrace k_{P}\rbrace)
    \ket{\sigma^{(1)},\cdots,\sigma^{(n)}},
\end{equation}
where $\sigma^{(j)}$ denote the spin of the particle whose position $x$ is the $j$-th smallest.
Let $\Pi^{(a,b)}$ be the operator that exchanges the $a$-th and $b$-th spins, i.e.,
\begin{equation}
\begin{split}
    &\Pi^{(a,b)}\ket{\sigma_1,\cdots,
    \sigma_a,\cdots,\sigma_b,\cdots,\sigma_n}\\
    &\equiv
    \ket{\sigma_1,\cdots,\sigma_b,
    \cdots,\sigma_a,\cdots,\sigma_n}.
\end{split}
\end{equation}
This operator satisfies
\begin{equation}
    \Pi^{(a,b)}\Pi^{(a,b)}=\mathrm{I}_{V_a\otimes V_b},\;\;\;\;
    \Pi^{(a,b)}\Pi^{(a,c)}\Pi^{(a,b)}=\Pi^{(b,c)}.
\end{equation}
We define
\begin{equation}
\label{eq_B:Y}
    Y^{(a,b)}(\xi)
    \equiv\frac{-ic'}{\xi+ic'}
    \mathrm{I}_{V_a\otimes V_b}
    +\frac{\xi}{\xi+ic'}\Pi^{(a,b)},
\end{equation}
and
\begin{equation}
    X^{(j,k)}(\xi)\equiv\Pi^{(j,k)}Y^{(j,k)}(\xi).
\end{equation}
Introducing the auxiliary space $\mathbb{C}^2_a$ with basis $\ket{\uparrow_a},\ket{\downarrow_a}$, we define the monodromy matrix
\begin{equation}
\begin{split}
    T^{(a)}(l)
    &=X^{(a,n)}(l-\lambda_n)\cdots X^{(a,1)}(l-\lambda_1)\\
    &\equiv\begin{pmatrix}
        A(l) & B(l)\\
        C(l) & D(l)
    \end{pmatrix},
\end{split}
\end{equation}
where $\lambda_j=k_{P_j}-ic'/2$, and the last expression is the matrix representation in the basis $\ket{\uparrow_a},\ket{\downarrow_a}$ of the auxiliary space $\mathbb{C}^2_a$.
Let $\sigma_1,\sigma_2,\sigma_3$ denote the single-particle Pauli matrices in the $x,y,z$ directions, respectively. For $u=1,2,3$, we define
\begin{equation}
\begin{split}
    \sigma^{(u)}_j
    &\equiv
    \mathrm{I}_{V_1}\otimes\cdots\otimes\mathrm{I}_{V_{j-1}}\otimes
    \sigma_u\otimes\mathrm{I}_{V_{j+1}}
    \otimes\cdots\otimes\mathrm{I}_{V_n}\\
    \sigma_j^\pm&\equiv\sigma_j^{(1)}
    \pm i\sigma_j^{(2)}.
\end{split}
\end{equation}
In the basis $\ket{\uparrow_a},\ket{\downarrow_a}$, $\Pi^{(a,j)}$ is represented as
\begin{equation}
    \Pi^{(a,j)}
    =\frac{1}{2}
    \begin{pmatrix}
        \mathbb{I}+\sigma^{(3)}_j & 2\sigma_j^-\\
        2\sigma_j^+ & \mathbb{I}-\sigma^{(3)}_j
    \end{pmatrix},
\end{equation}
which yields
\begin{equation}
\label{eq_B:X}
\begin{split}
    &X^{(a,j)}(\xi)\\
    &=\frac{1}{\xi+ic'}
    \begin{pmatrix}
        \left(\xi-\frac{ic'}{2}\right)\mathbb{I}-\frac{ic'\sigma^{(3)}_j}{2}&
        -ic'\sigma_j^-\\
        -ic'\sigma_j^+&
        \left(\xi-\frac{ic'}{2}\right)\mathbb{I}+\frac{ic'\sigma^{(3)}_j}{2}
    \end{pmatrix},
\end{split}
\end{equation}
where $\mathbb{I}\equiv \mathrm{I}_{V_1}\otimes\cdots\otimes\mathrm{I}_{V_n}$.
We define the reference (vacuum) state as
\begin{equation}
    \ket{\mathrm{vac}}
    \equiv
    \ket{\uparrow\cdots\uparrow}
    \in V_1\otimes\cdots\otimes V_n.
\end{equation}
Then,
\begin{equation}
\label{eq_B:A}
    A(l)\ket{\mathrm{vac}}
    =\prod_{j=1}^n\frac{l-\lambda_j-ic'}{l-\lambda_j+ic'}
    \ket{\mathrm{vac}},
\end{equation}
\begin{equation}
    C(l)\ket{\mathrm{vac}}
    =0,
\end{equation}
\begin{equation}
\label{eq_B:D}
    D(l)\ket{\mathrm{vac}}
    =\prod_{j=1}^n\frac{l-\lambda_j}{l-\lambda_j+ic'}
    \ket{\mathrm{vac}}.
\end{equation}
For the auxiliary spaces $\mathbb{C}^2_a$ and $\mathbb{C}^2_b$, the relation
\begin{equation}
    Y^{(a,b)}(l-\mu)T^{(a)}(l)
    T^{(b)}(\mu)
    =T^{(a)}(\mu)T^{(b)}(l)Y^{(a,b)}(l-\mu),
\end{equation}
holds, from which we obtain
\begin{equation}
    B(l)B(\mu)=B(\mu)B(l),
\end{equation}
\begin{equation}
\label{eq_B:AB}
    A(\mu)B(l)
    =\frac{l-\mu-ic'}{l-\mu}B(l)A(\mu)
    +\frac{ic'}{l-\mu}B(\mu)A(l),
\end{equation}
\begin{equation}
\label{eq_B:DB}
    D(\mu)B(l)
    =\frac{\mu-l-ic'}{\mu-l}B(l)D(\mu)
    +\frac{ic'}{\mu-l}B(\mu)D(l).
\end{equation}
Then Eqs.~\eqref{eq_B:bc} and \eqref{eq_B:pbc} are equivalent to the following expressions, respectively
\begin{equation}
\begin{split}
    &\Ket{k_{P_1},\cdots,k_{P_{j+1}},
    k_{P_j},\cdots,k_{P_n}}\\
    &=Y^{(j,j+1)}(k_{P_j}-k_{P_{j+1}})
    \Ket{k_{P_1},\cdots,k_{P_n}},
\end{split}
\end{equation}
\begin{equation}
\begin{split}
    &\Ket{k_{P_1},\cdots,k_{P_n}}\\
    &=-\exp[ik_{P_1}L]
    (A(\lambda_1)+D(\lambda_1))
    \Ket{k_{P_1},\cdots,k_{P_n}}.
\end{split}
\end{equation}
When the Bethe equations Eqs.~\eqref{eq_bethe_boson_1} and \eqref{eq_bethe_boson_2} are satisfied, we obtain
\begin{equation}
\begin{split}
    &B(l_1)\cdots B(l_M)\Ket{\mathrm{vac}}\\
    &=-\exp[ik_{P_1}L]
    (A(\lambda_1)+D(\lambda_1))
    B(l_1)\cdots B(l_M)\Ket{\mathrm{vac}}.
\end{split}
\end{equation}
Let $s(P)$ be a coefficient depending on the permutation $P$, and define
\begin{equation}
    \Ket{\Psi(P)}
    \equiv s(P)B(l_1)\cdots B(l_M)\Ket{\mathrm{vac}}.
\end{equation}
If this satisfies
\begin{equation}
    \Ket{\Psi(P\sigma_{j,j+1})}
    =Y^{(j,j+1)}(k_{P_j}-k_{P_{j+1}})
    \Ket{\Psi(P)},
\end{equation}
then the right eigenstate of $H_{eff}$ in Eq.~\eqref{eq:bethe_wavefunction} is given by

\begin{equation}
\begin{split}
    &\psi_R
    =\sum_{P\in\mathfrak{P}^n}
    \exp\left[
    i\sum_{j=1}k_{P_j}x_{Q_j}
    \right]
    \ket{A(Q,P)},\\
    &\ket{A(Q,P)}
    =\sum_{\sigma_1,\cdots,\sigma_n}\Braket{\sigma_{Q_1},\cdots,
    \sigma_{Q_n}
    |\Psi(P)}
    \Ket{\sigma_1,\cdots,\sigma_n}.
\end{split}
\end{equation}
For example, in the case $n=2$ and $M=1$, we have
\begin{equation}
    s(P)
    =\mathrm{sgn}(P)(k_{P_1}-k_{P_2}+ic'),
\end{equation}
and the eigenfunction is given by
\begin{equation}
\label{eq:bethe_wavefunction_2}
\begin{split}
    &\psi_R(x_1,x_2)\\
    &=A
    \lbrace
    \exp[i(k_1x_1+k_2x_2)]
    -\exp[i(k_2x_1+k_1x_2)]
    \rbrace\\
    &\;\;\;\;\times
    (\ket{\uparrow\downarrow}
    -\ket{\downarrow\uparrow}
    ),\\
    &A=
    \frac{-ic'(k_{1}-k_{2}+ic')(k_{2}-k_{1}+ic')}{(k_{1}-k_{2}+i3c')(k_{2}-k_{1}+i3c')}.
\end{split}
\end{equation}

\subsection{Highest-weight states}
We now show that $S^+\psi_R=0$. It suffices to prove that $S^+B(l_1)\cdots B(l_M)\Ket{\mathrm{vac}}=0$. From
\begin{equation}
    [X^{(a,j)}(\xi),I_a\otimes\sigma_j^{(u)}]
    =[\sigma_a^{(u)}\otimes\mathbb{I},X^{(a,j)}(\xi)],
\end{equation}
we obtain
\begin{equation}
    \label{eq_B:T_sigma}
    \left[T^{(a)}(l),\sum_{j=1}^n
    I_a\otimes\sigma_j^s\right]
    =[\sigma_a^s\otimes
    \mathbb{I},T^{(a)}(l)].
\end{equation}
Since
\begin{equation}
    S^+
    =\frac{1}{2}\sum_{j=1}^n
    \sigma_j^+
    =\frac{1}{2}\sum_{j=1}^n
    \left(\sigma_j^1\pm i\sigma_j^2
    \right),
\end{equation}
we have
\begin{equation}
    \frac{1}{2}\sum_{j=1}^n
    I_a\otimes\left(
    \sigma_j^{(1)}+ i\sigma_j^{(2)}
    \right)
    =\begin{pmatrix}
        S^+&0\\0&S^+
    \end{pmatrix},
\end{equation}
\begin{equation}
    \frac{1}{2}\left(
    \sigma_a^{(1)}+ i\sigma_a^{(2)}
    \right)
    \otimes\mathbf{I}_n
    =\begin{pmatrix}
        0&\mathbb{I}\\0&0
    \end{pmatrix}.
\end{equation}
From Eq.~\eqref{eq_B:T_sigma}, we obtain
\begin{equation}
\begin{split}
    &\left[T^{(a)}(l),\frac{1}{2}\sum_{j=1}^n
    \mathrm{I}_a\otimes\left(
    \sigma_j^{(1)}+ i\sigma_j^{(2)}
    \right)\right]\\
    &=\left[\frac{1}{2}\left(
    \sigma_a^{(1)}+ i\sigma_a^{(2)}
    \right)\otimes
    \mathbb{I},T^{(a)}(l)\right],
\end{split}
\end{equation}
which implies
\begin{equation}
    \begin{pmatrix}
        [A(l),S^+]&[B(l),S^+]\\
        [C(l),S^+]&[D(l),S^+]
    \end{pmatrix}
    =\begin{pmatrix}
        C(l)&D(l)-A(l)\\0&-C(l)
    \end{pmatrix},
\end{equation}
Hence,
\begin{equation}
    [S^+,B(l)]=A(l)-D(l).
\end{equation}
Thus,

\begin{widetext}
\begin{equation}
\begin{split}
    &S^+B(l_1)\cdots B(l_M)\ket{\mathrm{vac}}
    =\sum_{j=1}^M
    B(l_1)\cdots B(l_{j-1})
    (A(l_j)-D(l_j))
    B(l_{j+1})\cdots
    B(l_M)\ket{\mathrm{vac}}\\
    &=\sum_{j=1}^M
    M_j(l_1,\cdots,l_M)
    B(l_1)\cdots B(l_{j-1})
    B(l_{j+1})\cdots
    B(l_M)\ket{\mathrm{vac}},
\end{split}
\end{equation}
where, for $\alpha=1,\cdots,M$,
\begin{equation}
    M_\alpha
    =\prod_{\substack{\beta=1\\\beta\neq\alpha}}^M\frac{l_\beta-l_\alpha-ic'}{l_\beta-l_\alpha}
    \prod_{j=1}^n\frac{l_\alpha-\lambda_j-ic'}{l_\alpha-\lambda_j+ic'}
    -\prod_{\substack{\beta=1\\\beta\neq\alpha}}^M\frac{l_\alpha-l_\beta-ic'}{l_\alpha-l_\beta}
    \prod_{j=1}^n\frac{l_\alpha-\lambda_j}{l_\alpha-\lambda_j+ic'}.
\end{equation}
\end{widetext}
When the Bethe equations Eq.~\eqref{eq_bethe_boson_2} are satisfied, this vanishes. Therefore, the state constructed by the Bethe ansatz is a highest-weight state.

\subsection{The fermionic Yang-Gaudin model}
In this subsection, we describe only the points that differ from the bosonic case. For fermions, we have
\begin{equation}
    Y^{(a,b)}(\xi)
    =\frac{ic'}{\xi+ic'}
    \mathrm{I}_{V_a\otimes V_b}
    +\frac{\xi}{\xi+ic'}\Pi^{(a,b)},
\end{equation}
Accordingly,
\begin{equation}
\begin{split}
    &X^{(a,j)}(\xi)\\
    &=\frac{1}{\xi+ic'}
    \begin{pmatrix}
        \left(\xi+\frac{ic'}{2}\right)\mathbb{I}+\frac{ic'\sigma^{(3)}_j}{2}&
        ic'\sigma_j^-\\
        ic'\sigma_j^+&
        \left(\xi+\frac{ic'}{2}\right)\mathbb{I}-\frac{ic'\sigma^{(3)}_j}{2}
    \end{pmatrix}.
\end{split}
\end{equation}
In contrast to the bosonic case, we set $\lambda_j=k_{P_j}+ic'/2$. Then the properties of $A(l),B(l),C(l),D(l)$ are modified as follows
\begin{equation}
\begin{split}
    &A(l)\ket{\mathrm{vac}}=\ket{\mathrm{vac}},\\
    &C(l)\ket{\mathrm{vac}}=0,\\
    &D(l)\ket{\mathrm{vac}}
    =\prod_{j=1}^n\frac{l-\lambda_j}{l-\lambda_j+ic'}
    \ket{\mathrm{vac}},
\end{split}
\end{equation}
\begin{equation}
\begin{split}
    B(l)B(\mu)&=B(\mu)B(l),\\
    A(\mu)B(l)
    &=\frac{l-\mu+ic'}{l-\mu}B(l)A(\mu)
    -\frac{ic'}{l-\mu}B(\mu)A(l),\\
    D(\mu)B(l)
    &=\frac{\mu-l+ic'}{\mu-l}B(l)D(\mu)
    -\frac{ic'}{\mu-l}B(\mu)D(l).
\end{split}
\end{equation}
In this case as well, following the same procedure as in the bosonic case, one can show that $S^+\psi_R=0$ when the Bethe equations Eq.~\eqref{eq_bethe_fermion_2} are satisfied.

\section{ANALYSIS BASED ON THE SCHR\"{O}DINGER EQUATION}
In this Appendix, we analyze the properties of the two-body Yang–Gaudin model directly from the Schrödinger equation. The Hamiltonian of the two-body Yang–Gaudin model is given by
\begin{equation}
    H_{\mathrm{eff}}
    =-\frac{\hbar^2}{2m}\left(
    \frac{\partial^2}{\partial x_1^2}+\frac{\partial^2}{\partial x_2^2}
    \right)
    +2\left(c
    -\frac{i\hbar\gamma}{4}
    \right)
    \delta(x_1-x_2).
\end{equation}
Introducing the change of variables
\begin{equation}
    R=\frac{1}{2}(x_1+x_2),
    \;\;\;\;\;\;
    r=x_1-x_2,
\end{equation}
the Schrödinger equation becomes
\begin{equation}
    \left(-
    \frac{1}{2}\frac{\partial^2}{\partial R^2}-2\frac{\partial^2}{\partial r^2}
    +2c'\delta(r)
    -E
    \right)
    \Psi(R,r,\sigma_1,\sigma_2)
    =0.
\end{equation}
Since the Hamiltonian is independent of spin, the wave function can be written as a product of spatial and spin parts. Assuming separability in $R$ and $r$, we write $\Psi=\psi(R)\phi(r)\chi(\sigma_1,\sigma_2)$. Then $\psi(R)$ and $\phi(r)$ satisfy
\begin{equation}
    -\frac{1}{2\psi(R)}\frac{\partial^2 \psi(R)}{\partial R^2}=E_R,
\end{equation}
\begin{equation}
    -\frac{2}{\phi(r)}\frac{\partial^2 \phi(r)}{\partial r^2}
    +2c'\delta(r)
    =E_r,
\end{equation}
where $E=E_R+E_r$.
In the bosonic singlet case, $\phi(r)$ is an odd function and satisfies $\phi(0)=0$, so $\Psi$ vanishes at $x_1=x_2$. Therefore, the contact interaction does not contribute at $x_1=x_2$, and no dissipation occurs.

By contrast, in the bosonic triplet sector and fermionic singlet sector, $\phi(r)$ is an even function. Let $k\equiv\sqrt{-E_r/2}$. For $E_r\neq0$, corresponding to solutions obtained via the Bethe ansatz, the general solution for the relative coordinate in the region $0\leq r\leq L$ is
\begin{equation}
    \phi(r)
    =Ae^{kr}+Be^{-kr}.
\end{equation}
Since $\phi(r)$ is even and $k\neq -k$, we have $A=B$. Thus $\phi(0)=2A\neq0$, implying that in the bosonic triplet sector and fermionic singlet sector, the contact interaction at $x_1=x_2$ gives rise to dissipation.

\end{document}